%% file: CFE_Carry_Trade_Copula_Paper_MA_14_Nov_2013_v3_0_0.tex
\documentclass[a4paper,11pt]{article}

\usepackage[square,sort]{natbib}

\usepackage{graphicx,verbatim,array,multicol}
\usepackage{amssymb}
\usepackage{amsfonts}
\usepackage{amsmath}
\usepackage{epsfig}
\usepackage{latexsym}
\usepackage{amssymb}
\usepackage{color}
\usepackage{amscd}
\usepackage{multirow}
\usepackage{graphicx}
\usepackage{lscape}
\usepackage{bm}

\usepackage{booktabs}
\usepackage{algpseudocode}
\usepackage{algorithm}
\usepackage{epstopdf} 
\usepackage{morefloats}
\usepackage[margin=1cm]{caption}

\usepackage{placeins}
\usepackage{tablefootnote}

\usepackage{pdflscape}

    \usepackage[breaklinks = true,linktocpage,
				pagebackref,colorlinks = true,
                 linkcolor = blue,
                 urlcolor  = blue,
                 citecolor = blue,
                 hyperindex = true,
                 hyperfigures = true
                 ]{hyperref}

\renewcommand{\baselinestretch}{1.5}

\newcommand{\Remm}[1]{}
\newtheorem{theo}{Theorem}[section]

\newtheorem{remark}[theo]{Remark}

\newtheorem{model ass}[theo]{Model Assumptions}
\newtheorem{defi}[theo]{Definition}

\numberwithin{equation}{section}

\csname @openrightfalse\endcsname

\makeatletter
\def\@author#1{\g@addto@macro\elsauthors{\normalsize%
    \def\baselinestretch{1}%
    \upshape\authorsep#1\unskip\textsuperscript{%
      \ifx\@fnmark\@empty\else\unskip\sep\@fnmark\let\sep=,\fi
      \ifx\@corref\@empty\else\unskip\sep\@corref\let\sep=,\fi
      }%
    \def\authorsep{\unskip,\space}%
    \global\let\@fnmark\@empty
    \global\let\@corref\@empty  
    \global\let\sep\@empty}%
    \@eadauthor={#1}
}
\makeatother

\begin{document}

\begin{center}{\large {\textbf{ Reinvestigating the Uncovered Interest Rate Parity Puzzle via Analysis of Multivariate Tail Dependence in Currency Carry Trades}}}
\end{center}

\vspace{0.2cm}

\begin{center}
\textbf{Matthew Ames} (Corresponding author\footnote{email: m.ames.12@ucl.ac.uk}) \\
{\footnotesize Department of Statistical Science, University College London, UK;}
\\
\vspace{5mm}

\textbf{Guillaume Bagnarosa} \\
{\footnotesize Director of Research, Molinero Capital Management, London, UK;\\
Honorary Research Associate, Department of Computer Science, University College London, UK}
\\
\vspace{5mm}

\textbf{Gareth W. Peters} \\
{\footnotesize Lecturer, Department of Statistical Science, University College London, UK;\\
Adjunct Scientist, CMIS, Commonwealth Scientific and Industrial Research Organisation, Australia; \\
Research Associate, Oxford-Man Institute, Oxford University, UK}

\end{center}

\begin{abstract}
\noindent
The currency carry trade is the investment strategy that involves selling low interest rate currencies in order to purchase higher interest rate currencies, thus profiting from the interest rate differentials. This is a well known financial puzzle to explain, since assuming foreign exchange risk is uninhibited and the markets have rational risk-neutral investors, then one would not expect profits from such strategies. That is uncovered interest rate parity (UIP), the parity condition in which exposure to foreign exchange risk, with unanticipated changes in exchange rates, should result in an outcome that changes in the exchange rate should offset the potential to profit from such interest rate differentials. The two primary assumptions required for interest rate parity are related to capital mobility and perfect substitutability of domestic and foreign assets. Given foreign exchange market equilibrium, the interest rate parity condition implies that the expected return on domestic assets will equal the exchange rate-adjusted expected return on foreign currency assets.

However, it has been shown empirically, that investors can actually earn arbitrage profits by borrowing in a country with a lower interest rate, exchanging for foreign currency, and investing in a foreign country with a higher interest rate, whilst allowing for any losses (or gains) from exchanging back to their domestic currency at maturity. Therefore trading strategies that aim to exploit the interest rate differentials can be profitable on average. The intention of this paper is therefore to reinterpret the currency carry trade puzzle in light of heavy tailed marginal models coupled with multivariate tail dependence features in the analysis of the risk-reward for the currency portfolios with high interest rate differentials and low interest rate differentials. To achieve this analysis of the multivariate extreme tail dependence we develop several parametric models and perform detailed model comparison. 

\vspace{0.5cm} \noindent \textbf{Keywords:} Currency carry trade, Multivariate tail dependence, Forward premium puzzle, Mixture models, Generalized Archimedean copula and Extreme value copula.
\end{abstract}

\newpage

\renewcommand{\arraystretch}{2}

\input{Introduction_MA_14_Nov_2013_v3_0_0}

\input{CurrencyCarryTrade_MA_14_Nov_2013_v3_0_0}

\input{DataDescriptionBasketFormation_MA_14_Nov_2013_v3_0_0}

\input{CopulaModels_MA_14_Nov_2013_v3_0_0}

\input{FittingtheModels_MA_14_Nov_2013_v3_0_0}

\input{Joint_Tail_Risk_Exposure_MA_28_Nov_2013_v4_0_0}

\input{Results_MA_14_Nov_2013_v3_0_0}
\input{Discussion_MA_14_Nov_2013_v3_0_0}


\bibliographystyle{unsrtnat}
\bibliography{CFE_COPULA_Bibliography}

\newpage

\input{appendixA_MA_14_Nov_2013_v3_0_0}

\input{appendixB_MA_14_Nov_2013_v3_0_0}

\input{appendixC_MA_06_Dec_2013_v3_0_0}

\end{document}

%% file: Introduction_MA_14_Nov_2013_v3_0_0.tex
\section{Introduction}

Understanding the behaviour of currency markets has been an active area of research for the past few decades. Much of the literature has focused on the marginal behaviours of exchange rates and portfolio returns from carry trade violations of UIP. However, there have been fewer studies investigating the joint exchange rate behaviours for the currency baskets used in a carry trade and the effect this may have on portfolio risk. One of the most robust puzzles in finance still to be satisfactorily explained is the uncovered interest rate parity puzzle and the associated excess average returns of currency carry trade strategies. Such trading strategies are popular approaches which involve constructing portfolios by selling low interest rate currencies in order to buy higher interest rate currencies, thus profiting from the interest rate differentials. When such portfolios are highly leveraged this can result in sizeable profits. The presence of such opportunities, pointed out by \citet{hansen1980,fama1984,backus2001affine} and more recently by \citet{Lustig2007,Brunnermeier2008,burnside2011peso,christiansen2011time,Lustig2011,Menkhoff2012}, violates the fundamental relationship of uncovered interest rate parity (UIP). The UIP refers to the parity condition in which exposure to foreign exchange risk, with unanticipated changes in exchange rates, is uninhibited and therefore if one assumes rational risk-neutral investors, then changes in the exchange rates should offset the potential to profit from the interest rate differentials between high interest rate (investment) currencies and low interest rate (funding) currencies. The existence of a working UIP relationship is related to two primary assumptions which are the capital mobility and the perfect substitutability of domestic and foreign assets. When UIP holds, then given foreign exchange market equilibrium, the interest rate parity condition implies that the expected return on domestic assets will equal the exchange rate-adjusted expected return on foreign currency assets. Therefore, no arbitrage opportunities should arise in practice; however such opportunities are routinely observed and exploited by large volume trading strategies.


Among the justifications of this phenomenon, \cite{fama1984} initially proposed a time varying risk premium within the forward rate relative to the associated spot rate. Thus, carry trade returns are viewed as compensation for higher risk exposure during recession and crisis periods.
Following this seminal paper, \cite{Lustig2011} employ two risk factors to price the cross section of carry trade
returns. The first risk factor is a level factor, named ``dollar risk factor'', which is essentially the average excess return on all foreign currency
portfolios. The second risk factor is a “slope” factor in exchange rates, named $HML_{FX}$. High interest rate currencies load more on this slope factor than low interest rate currencies. This factor is found to account for most of the cross-sectional variation in average excess returns between high and low interest rate currencies.
\cite{Menkhoff2012} consider a related approach using innovations in global FX volatility in place of the $HML_{FX}$ factor. The authors demonstrate that high interest rate currencies tend to be negatively related to the innovations in global FX volatility which is considered as a proxy for unexpected changes in the FX market volatility. It is shown that global FX volatility is a pervasive risk factor in the cross section of FX excess returns and that its pricing power extends to several other test assets. Liquidity is also shown to be a useful risk factor in understanding carry trade returns, but to a lesser extent than global FX volatility.

Another hypothesis, proposed by \cite{Farhi2008}, consists in justifying this puzzle through the inclusion of a mean reverting risk premium. According to their model a risky country, which is more sensitive to economic extreme events, represents a high risk of currency depreciation and has thus to propose, in order to compensate this risk, a higher interest rate. Then, when the risk premium reverts to the mean, their exchange rate appreciates while they still have a high interest rate which thus replicates the forward rate premium puzzle. 
Furthermore, \cite{Weitzman2007} demonstrates through a Bayesian approach that the uncertainty about the variance of the future growth rates combined with a thin-tailed prior distribution would generate the fat-tailed distribution required to solve the forward premium puzzle.

The causality relation between the interest rate differential and the currency shocks can be presented the other way around as detailed in \cite{Brunnermeier2008}. In this article, the authors indeed assume that the currency carry trade mechanically attracts investors and more specifically speculators who accordingly increase the probability of a market crash. Tail events among currencies would thus be caused by speculators' need to unwind their positions when they get closer to funding constraints.


\vspace{5mm}


In this paper we build on the existing literature by studying a stochastic feature of the joint tail behaviours of the currencies within each of the long and the short baskets, which form the carry trade. We aim to explore to what extent one can attribute the excess average returns with regard to compensation for exposure to tail risk, for example either dramatic depreciations in the value of the high interest rate currencies or dramatic appreciations in the value of the low interest rate currencies in times of high market volatility.
We postulate that such analyses should also benefit from consideration not only of the marginal behaviours of the processes under study, in this case the exchange rates of currencies in a portfolio, but also a rigorous analysis of the joint dependence features of such relationships. Therefore, we investigate such joint relationships in light of the UIP condition. To achieve this, we study the probability of joint extreme movements in the funding and investment currency baskets and interpret these extremal tail probabilities as relative risk exposures of adverse and beneficial joint currency movements which would affect the portfolio returns. In addition this allows us to obtain a relative contribution to the exposure of the portfolio profit decomposed in terms of the downside and upside risks that are contributed from such tail dependence features in each currency basket. We argue that the analysis of the carry trade is better informed when the model analysis is performed jointly by modelling the multivariate behaviour of the marginal processes of currency baskets accounting for potential multivariate extremes, and multivariate tail dependence features, whilst still incorporating heavy-tailed relationships studied in marginal processes. To achieve the accurate modelling of the potentially complex multivariate dependence features the statistical models considered should be sufficiently flexible to accommodate a variety of potential tail dependence relationships. In this regard we consider two flexible families of mixture copula models comprised of members of the Archimedean copula family as well as the outer-power Clayton copula family, all of which admit different degrees of asymmetric tail dependence.

The approach we adopt in this paper is a statistical framework with a high degree of sophistication, however its fundamental reasoning and justification is indeed analogous in nature to the ideas considered when investigating the ``equity risk premium puzzle" coined by \cite{Mehra1985} in the late 80's. The equity risk premium puzzle effectively refers to the fact that demand for government bonds which have lower returns than stocks still exists and generally remains high. This poses a puzzle for economists to explain why the magnitude of the disparity between the returns on each of these asset classes, stocks versus bonds, known as the equity risk premium, is so great and therefore implies an implausibly high level of investor risk aversion. In the seminal paper written by \cite{Rietz1988}, the author proposes to explain the ``equity risk premium puzzle" \cite{Mehra1985} by taking into consideration the low but still significant probability of a joint catastrophic event. 

Analogously in this manuscript, we are proposing to explore the highly leveraged opportunities in currency carry trades that arise due to violation of the UIP. However, we conjecture that if the assessment of the risk associated with such trading strategies was modified to adequately take into account the potential for joint catastrophic risk events accounting for the non-trivial probabilities of joint adverse movements in currency exchange rates, then such strategies may not seem so profitable relative to the risk borne by the investor. We propose a rigorous probabilistic model in order to quantify this phenomenon and potentially detect when liquidity in FX markets may dry up. This probabilistic measure of dependence can then be very useful for risk management of such portfolios but also for making more tractable the valuation of structured products or other derivatives indexed on this specific strategy. To be more specific, the principal contribution of our article is indeed to model the dependencies between exchange rate log returns using flexible families of mixture copula models comprised of members of the Archimedean copula family as well as the outer-power Clayton copula family. This probabilistic approach allows us to express the joint distribution of the vectors of random variables, in our case vectors of exchange rate log-returns in each basket of currencies, as functions of each marginal distribution and the copula function itself. 

Whereas in the literature we mentioned earlier, the tail thickness resulting from the carry trade has been either treated individually for each exchange rate or through the measurement of distribution moments that may not be adapted to a proper estimation of the extremal tail dependencies, such as covariance. In our paper, we propose instead to build on a daily basis a set of baskets of currencies with regards to the interest rate differentials of each currency with the US dollar. Using a mixture of copula functions, we then extract a measure of the tail dependencies within each basket and finally interpret the results. Among the outcomes of our study, we demonstrate that during the crisis periods, the high interest rate currencies tend to display very significant upper tail dependence. Accordingly, we thus conclude that the appealing high return profile of a carry portfolio is not only compensating the tail thickness of each individual component probability distribution but also the fact that they tend to occur simultaneously and lead to a portfolio particularly sensitive to the risk of drawdown. Furthermore, we also demonstrate that high interest rate currency baskets and low interest rate currency baskets can display periods during which the tail dependence gets inverted, demonstrating when periods of construction of the aforementioned carry positions are being undertaken by investors.

Therefore, our analysis provides interesting insight into the seemingly complex understanding of how on average profits are regularly made on highly leveraged portfolios, however the presence of multivariate tail dependence features in the assets of the carry baskets will change the manner in which one should assess the risk associated with such highly leveraged profits. This may help to explain the violation of the UIP condition when such risk analysis is appropriately considered for the joint behaviour. It is our conjecture that such profits and the risk metrics considered when assessing the risk-return relationships of such trading strategies should be therefore altered to account for tail dependence features in the baskets constructed from interest rate differentials. We would therefore argue that failure to take into consideration the demonstrated potential for significant losses associated with a non-trivial joint probability of large adverse movements in exchange rates could wipe out all potential average profits should such an extreme event occur.






The paper is structured as follows. In the second section, we review the concept of UIP and the associated carry trade. The third section presents details of the portfolio construction as well as a detailed description of the data utilised in the analysis.  The fourth section is devoted to the description of the copula models and their applications in finance. We also describe the model we propose to measure the tail dependence within the high interest rate and low interest rate baskets. Section five presents estimation of the multivariate distributions for the mixture copula and marginal heavy tailed models via the Inference Function for the Margins approach. Section six examines the joint tail exposure in currency carry trades via an analysis of tail dependence. Section seven presents results and comments about the carry trade portfolio returns when we take into account the tail dependence exposure. Finally section eight provides a discussion of our findings in this paper.

%% file: CurrencyCarryTrade_MA_14_Nov_2013_v3_0_0.tex
\section{Currency Carry Trade Problem Formulation}
\label{section:problem}

\subsection{The Uncovered Interest Rate Parity Puzzle}

In this section we define the concept of uncovered interest rate parity and the associated assumptions which lead to the economic theory that one should not be able to profit from the carry trade strategy.
The Uncovered interest rate parity (UIP) condition is directly linked to the arbitrage relation existing between the spot and the forward prices of a given currency, namely the Covered interest rate parity (CIP) condition.

\begin{defi}
Covered Interest Rate Parity (CIP) \\
This relation states that the price of a forward rate can be expressed as follows:
\begin{align} 
F_{t}^{T}=e^{(r_{t}-r_{t}^{f})(T-t)}S_{t} \label{Forward_Pricing}
\end{align}
where $F_{t}^{T}$ and $S_{t}$ denote respectively the forward and the spot prices at time $t$. While $r_{t}$ and $r_{t}^{f}$ represent the local risk free rate\footnote{We mean by local risk free rate the interest rate prevailing in the reference country which would be for instance the dollar for an American investor.} and the foreign risk free rate. We denote by $T$ the maturity of the forward contract considered.
The CIP condition states that one should not be able to make a risk free profit by selling a forward contract and replicating its payoff through the spot market.
\end{defi}

It is worth emphasizing that under the hypothesis of absence of arbitrage opportunities, the CIP relation is directly resulting from the replication of the forward contract payoff using a self financed strategy. Moreover, the validity of this arbitrage relation has been demonstrated empirically in the currency market \citet{Juhl2006,Akram2008} when one considers daily data. The highly unusual period following the onset of the financial crisis in August 2007 saw large deviations from CIP due to the funding constraints of arbitrageurs and uncertainty about counterparty risk, see \cite{coffey2009capital} for a thorough analysis of this period, though this was an exceptional case and typically CIP holds.

%
%
%

\begin{defi}
Uncovered Interest Rate Parity (UIP) \\
If we assume the forward price is a martingale under the risk neutral probability $\mathbb{Q}$ \cite{musiela2011martingale} then the fair value of the forward contract at time $t$ equals:
\begin{align}
E_{\mathbb{Q}}[ S_{T}|\mathcal{F}_{t}]= F_{t}^{T} \label{martingale}
\end{align}
where $\mathcal{F}_{t}$ is the filtration associated to the stochastic process $S_{t}$. Replacing the expression (\ref{martingale}) in the relation (\ref{Forward_Pricing}) leads to the UIP equation:
\begin{align}
E_{\mathbb{Q}}\Bigg[\frac{S_{T}}{S_{t}}\Bigg|\mathcal{F}_{t}\Bigg]=\frac{F_{t}^{T}}{S_{t}}=e^{(r_{t}-r_{t}^{f})(T-t)} \label{UIP}
\end{align}
The UIP equation states that under the risk neutral probability the expected variation of the exchange rate $S_{t}$ should equal the differential of interest rate between the two countries. Thus if an investor borrows $S_t e^{-r^{f}_{t} (T-t)}$, converts it into foreign currency and invests it in the foreign risk free bond and finally converts it back. The profit or the loss resulting at the maturity date T should be equal to $S_{t}e^{(r_{t}-r_{t}^{f})(T-t)}$ which was the price paid initially for the forward contract under the hypothesis of absence of arbitrage opportunities.\\ 
\end{defi}

Unlike the CIP condition, where we hedge the exchange rate risk by selling a forward contract, it is not true that UIP regularly holds in practice. Numerous empirical studies \citet{fama1984,hansen1980,engel1996forward,Lustig2007} have previously demonstrated the failure of UIP, i.e. that investors can actually earn profits by borrowing in a country with a lower interest rate, exchanging for foreign currency, and investing in a foreign country with a higher interest rate, whilst allowing for any losses (or gains) from exchanging back to their domestic currency at maturity. Therefore, trading strategies that aim to exploit the interest rate differentials can be profitable on average. 
This is notably the case for the currency carry trade which is thus the simple investment strategy of selling a high interest rate currency forward and then buying a low interest rate currency forward. The idea is that the interest rate returns will outweigh any potential adverse moves in the exchange rate. Historically the Japanese yen and Swiss franc have been used as ``funding currencies", since they have maintained very low interest rates for a long period. The currencies of developing nations, such as the South African rand and Brazilian real have been typically used as ``investment currencies". Whilst this sounds like an easy money making strategy there is of course a downside risk. This risk comes in the form of currency crashes in periods of global FX volatility and liquidity shortages. A prime example of this is the sharp yen carry trade reversal in 2007.

\subsection{Problem Statement}

This paper explores the non-linear stochastic dependence structures of the currency baskets constructed in carry trade strategies and then interprets how and why these features lead to the average profitability observed empirically in carry trades being exposed to significant extreme tail risk. The standard mean-variance framework is not able to capture the complete set of risk exposures being taken in such strategies. In order to model the risk present in currency carry strategies one not only needs to consider heavy tails for the distributions of individual exchange rate log returns, but also the joint tail dependence between the currencies in the baskets. That is to say standard portfolio optimisation focusing on mean-variance neglects the very important risk of large joint extremal adverse movements in the exchange rates of the currencies in the carry trade baskets, that is captured in tail risk as measured by a notion of concordance for extreme events which can be significant even in cases where correlation is insignificant between currencies in the basket.

%% file: DataDescriptionBasketFormation_MA_14_Nov_2013_v3_0_0.tex
\section{Exchange Rate Multivariate Data Description and Currency Portfolio Construction}

We initially consider for our empirical analysis a set of 49 currency exchange rates relative to the USD, as in \cite{Menkhoff2012}.
Since the aim of this investigation is to analyse the tail dependence present in the long and short baskets of a carry trade, we removed a number of currencies from consideration (either entirely or for certain periods). Most notably, we entirely removed the euro area countries from consideration since throughout the period 1989 to 1999 these currencies were tightly banded by the ERM in preparation for monetary unification. Thus the characteristics of the returns from these currencies were heavily influenced by a so-called `convergence trade' to the single currency.
A number of other currencies were pegged to the US dollar and thus the interest rate differentials calculated from market data can be misleading as the forward price is tightly banded, i.e. the interest rate differential is noisy. The following currencies were pegged to the US dollar for all or some of the sample period and hence were removed from consideration for that period: Hong Kong dollar (HKD), Kuwait dollar (KWD), Malaysian ringgit (MYR) and Saudia Arabian riyal (SAR).
Also, the Danish krone which was pegged to the German mark and then later to the euro was removed for the same reason.

We indeed considered the point of view of an American investor as this is generally the hypothesis retained in the literature \citet{Brunnermeier2008, Menkhoff2012}. However the same analysis could be carried out from any other investor standpoint as the phenomena we will describe does not only depend on a specific currency but on two or more sets of currencies. These sets of currencies correspond to the high interest rate currencies which are used to obtain the highest return (named the ``investment currencies") and the low interest rate currencies which allows for borrowing at a low cost the amount of money necessary for this investment (named the ``financing currencies").

The time series analysed range from 02/01/1989 to 31/05/2013 and comprise the following 33 currencies: 
Australia (AUD), Brazil (BRL), Canada (CAD), Croatia (HRK), Cyprus (CYP), Czech Republic (CZK), Euro area (EUR), Greece (GRD), Hungary (HUF), Iceland (ISK), India (INR), Indonesia (IDR), Israel (ILS),
Japan (JPY), Malaysia (MYR), Mexico (MXN), New Zealand (NZD), Norway (NOK), Philippines (PHP), Poland (PLN), Russia (RUB), Singapore (SGD), Slovakia (SKK), Slovenia (SIT), South Africa (ZAR), South Korea (KRW), Sweden (SEK), Switzerland (CHF), Taiwan (TWD), Thailand (THB), Turkey (TRY), Ukraine (UAH) and the United Kingdom (GBP).


We have considered daily settlement prices for each currency exchange rate as well as the daily settlement price for the associated 1 month forward contract. We utilise the same dataset (albeit starting in 1989 rather than 1983) as studied in \cite{Lustig2011} and \cite{Menkhoff2012} in order to replicate their portfolio returns without tail dependence risk adjustments. Due to differing market closing days, e.g. national holidays, there was missing data for a couple of currencies and for a small number of days. For missing prices, the previous day's closing prices were retained.

We consider two views on the analysis of the dataset. Firstly, we consider a daily analysis since we have at our disposal one observation per currency per day which makes our analysis of individual tails and their interdependencies more robust. We then consider a monthly analysis, since in order to build a carry portfolio the position has to be held until the maturity of the forward contract, which in our case is a 1 month horizon, and hence one observation is retained per currency at the end of each month.



 Among the currencies under scrutiny, some of them have displayed very large variations in the last decade mainly for macro-economic reasons. Therefore, we considered it insightful to mention some of the most meaningful. The Brazilian real displays in its time series two important periods of shocks, the first in 2001 and the second in 2002. Naturally the first of them was due to the terrorist attacks against the world trade center in September. However the Brazilian real has been also impacted by the market's concerns of a contagion after the rumours of default of the Argentinian government. The second shock on the Brazilian real in 2002 was related to the potential election of the Workers' Party leader Luiz Inacio Lula da Silva which prompted concern he might spark a default by overspending to meet promises of spurring growth and employment. In 2001, the South African rand slumped 29\% after the events of September 11 and the market's concern of a global recession leading to a slump in commodity prices to which the South African economy is particularly exposed to. As a third example of a shock in an instrumental currency in a carry trade strategy we note the 30\% daily loss of the Turkish lira on the 22nd of February 2001. This was due to Turkey's decision to abandon the defence of their currency in order to reduce the cost of financing lira-denominated debt. It is worth mentioning that we did not remove these data points from our time series given that different events may have impacted the other exchange rates at a different time but our analysis does not focus only on the tail events associated to a particular currency but more on the events impacting simultaneously a set of currencies.

\subsection{ Currency Portfolios Formation}

As described earlier, the currency carry trade results from the differential of interest rates prevailing in different countries. By borrowing a certain amount of money in low interest rate countries and investing it in high interest rate countries, a recurrent profit can be generated given that the UIP condition is on average not satisfied. In order to differentiate the ``financing currencies" from the ``investment currencies", we start by classifying each currency relative to its differential of risk free rate with the US dollar.  We note the following basic explanation of the high rates and low rates. In general, countries that are considered `safe' can borrow at a lower interest rate, which may explain why historically the US dollar or Swiss franc interest rates were low \cite{Gourinchas2007} while the Turkish lira rates were historically high as this country is not considered as financially secure. 
Moreover we demonstrated in expression (\ref{UIP}) that the differential of interest rates between two countries can be estimated through the ratio of the forward contract price and the spot price, see \cite{Juhl2006} who show this holds empirically on a daily basis. Accordingly, instead of considering the differential of risk free rates between the reference and the foreign countries, we build our respective baskets of currencies with respect to the ratio of the forward and the spot prices for each currency. On a daily basis we compute this ratio for each of the $n$ currencies (available in the dataset on that day) and then build five baskets. The first basket gathers the $n/5$ currencies with the highest positive differential of interest rate with the US dollar. The selected currencies over the period 02/01/1989 to 31/05/2013 for the high interest rate basket are displayed in Figure \ref{b5_49} [Appendix A]. 
These currencies are thus representing the ``investment" currencies, through which we invest the money to benefit from the currency carry trade. The last basket will gather the $n/5$ currencies with the highest negative differential (or at least the lowest differential) of interest rate. We display the low interest rate currency selections in Figure \ref{b1_49} [Appendix A].
These currencies are thus representing the ``financing" currencies, through which we borrow the money to build the currency carry trade.


Conditionally to this classification we investigate then the joint distribution of each group of currencies to understand the impact of the currency carry trade, embodied by the differential of interest rates, on currencies returns. In our analysis we concentrate on the high interest rate basket (investment currencies) and the low interest rate basket (funding currencies), since typically when implementing a carry trade strategy one would go short the low interest rate basket and go long the high interest rate basket.

%
%
%
%
%
%
%
%

%% file: CopulaModels_MA_14_Nov_2013_v3_0_0.tex
\section{Copula Modelling for Currency Exchange Rate Baskets}

In order to capture asymmetries in the upper and lower tail dependence in each currency basket we consider three models; two Archimedean mixture models and the outer power transformed Clayton copula. The mixture models considered are the Clayton-Gumbel mixture and the Clayton-Frank-Gumbel mixture, where the Frank component allows for periods of no tail dependence within the basket as well as negative dependence. To be clear, we fit these copula models to each of the long and short baskets separately. Typically the number of currencies in each of the long basket (investment currencies) and the short basket (funding currencies) is 4 or 5, however this occasionally varies across the time period considered from a minimum of 2 up to a maximum of 6.


In this section we consider two stages, firstly the estimation of suitable heavy tailed marginal models for the currency exchange rates (relative to USD), followed by the estimation of the dependence structure of the multivariate model composed of multiple exchange rates in currency baskets for long and short positions. To achieve this we consider a copula model framework where we utilise the well-known result of Sklar's theorem (see Theorem \ref{eq:sklar}) as it allows one to separate the multivariate distribution into its marginal distributions and the dependence structure between the margins known as a copula distribution, which is unique for continuous marginal models.

%
%

\begin{theo} Sklar's Theorem (1959) \\
Consider a d-dimensional cdf $F$ with marginals $F_1 , \ldots , F_d$. There exists a copula C, s.t.
\begin{equation}
F(x_1,...,x_d) = C(F_1(x_1),\ldots,F_d(x_d))
\label{eq:sklar}
\end{equation}
for all $x_i \in (-\infty, \infty) , i \in 1,\ldots,d$.
\noindent Furthermore, if $F_i$ is continuous for all $i = 1,\ldots, d$ then C is unique;
otherwise C is uniquely determined only on $Ran F_1 \times \cdots \times Ran F_d$, where $Ran F_i$ denotes the
range of the cdf $F_i$. 
\end{theo}

This result demonstrates how copula models therefore provide a mechanism to model the marginal behaviour of each currency and then separately to focus on developing hypotheses regarding the possible dependence structures between the log returns of the forward exchange rates of the currencies in the baskets, which can be tested through parameterization of a model via a copula and then a process of model selection.

\subsection{{\bf Archimedean Mixture Copula Models}}

There are many possible copula models that could be considered in the modelling of the multivariate dependence features of the currency baskets. 
Previously in the exchange rate modelling literature \cite{patton2006modelling} considered a symmetrized Joe-Clayton copula in order to capture the asymmetric dependence structure between the Deutsche mark and the yen. The author demonstrated the time-varying nature of the dependence, noting that the mark-dollar and yen-dollar exchange rates were more correlated when they were depreciating against the dollar than when they were appreciating. A flexible time-varying copula model is introduced by \cite{dias2010modeling} using the Fisher information to specify the dynamic correlation. The authors find a significantly time-varying correlation between Euro/US dollar and Yen/US dollar, dependent on the past return realizations. However, the UIP puzzle introduced in Section \ref{section:problem} has not been explored in the literature from a copula modelling approach.
The intention of this analysis was to work with models that have well understood tail dependence features and are relatively parsimonious with regard to the number of parameters specifying the copula and the resulting tail dependence. 



In order to add additional flexibility in the possible dependence features one can study for the currency baskets, we decided to utilize mixtures of copula models. In this regard we have the advantage that we can consider asymmetric dependence relationships in the upper tails and the lower tails in the multivariate model. In addition we can perform a type of model selection purely by incorporating into the estimation the mixture weights associated with each dependence hypothesis. That is the data can be utilised to decide the strength of each dependence feature as interpreted directly through the estimated mixture weight attributed to the feature encoded in the particular mixture component from the Archimedean family.

In particular we have noted that mixture copulae can be used to model asymmetric tail dependence, i.e. by combining the one-parameter families discussed in Section \ref{section:oneparam} or indeed by any combination of copulae. This is possible since a linear convex combination of 2 copulae is itself a copula, see \cite{nelsen2006}.

\begin{defi}
Mixture Copula \\
A mixture copula is a linear weighted combination of copulae of the form:
\begin{equation}
C_M({\bf u};{\bf \Theta}) = \sum_{i=1}^N \lambda_i C_i ({\bf u};{\bf \theta_i}) 
\end{equation}
where  $0 \leq \lambda_i \leq 1 \;\; \forall i = 1, ..., N$ and $\sum_{i=1}^N \lambda_i = 1$
\end{defi}

Thus we can combine a copula with lower tail dependence, a copula with positive or negative dependence and a copula with upper tail dependence to produce a more flexible copula capable of modelling the multivariate log returns of forward exchange rates of a basket of currencies. For this reason in this analysis we will use the Clayton-Frank-Gumbel mixture model. In addition to the C-F-G mixture model we will also investigate the C-G mixture model and the outer power Clayton copula model.

\begin{remark} We note that the tail dependence of a mixture copula can be obtained as the linear weighted combination of the tail dependence of each component in the mixture weighted by the appropriate mixture weight, as discussed in for example \cite{nelsen2006} and \cite{peters2012copula}.
\end{remark}


A function $\psi$ is said to generate an Archimedean copula if it satisfies the properties in Definition \ref{archm_gen}.

\begin{defi} Archimedean Generator \\
An Archimedean generator is a continuous, decreasing function $\psi:[0, \infty) \rightarrow [0, 1]$ which satisfies the following conditions:
\begin{enumerate}
\item $\psi(0) = 1$
\item $\psi(\infty) = lim_{t \rightarrow \infty} \psi(t) = 0$
\item $\psi$ is strictly decreasing on $[0, inf\{t: \psi(t) = 0\}]$
\end{enumerate}
\label{archm_gen}
\end{defi}

\begin{defi} Archimedean Copula\\
A d-dimensional copula C is called Archimedean if for some generator $\psi$ it can be represented as:
\begin{equation}
C({\bf u}) = \psi \{\psi^{-1}(u_1) + \cdots + \psi^{-1}(u_d)\} = \psi \{t({\bf u}) \} \;\;\;\; \forall {\bf u} \in [0, 1]^d
\label{eq:archimedean_copula}
\end{equation}
where $\psi ^{-1}:[0, 1] \rightarrow [0, \infty)$ is the inverse generator with $\psi^{-1}(0) = inf\{t: \psi(t) = 0\}$.
\end{defi}

\noindent Note the shorthand notation $t({\bf u}) = \psi^{-1}(u_1) + \cdots + \psi^{-1}(u_d)$ that will be used throughout this section.

\subsubsection{Finding Archimedean Copula Densities}

As we will see later, it is necessary to have formulas for computing the copula densities if one seeks to fit these models using a maximum likelihood approach. Equation~\ref{eq:density} provides such a formula in a generic form for each member of the family of Archimedean copulae.

\vspace{3mm}

\begin{defi} Archimedean Copula Density \\
\cite{mcneil2009} prove that an Archimedean copula C admits a density c if and only if $\psi^{(d-1)}$ exists and is absolutely continuous on $(0, \infty)$. When this condition is satisfied, the copula density c is given by
\begin{equation}
c({\bf u})\;\; = \;\; \frac{ \partial^d C(u_1, \ldots , u_d)}{\partial u_1 \ldots \partial u_d} \;\;
= \;\;\ \psi^{(d)} \{t({\bf u})\} \prod_{j=1}^d (\psi^{-1})'(u_j) \;\; , \;\;\;\; {\bf u} \in (0, 1)^d
\label{eq:density}
\end{equation}
\end{defi}


As has been noted above, in performing the estimation of these transformed copula models via likelihood based inference it will be of great benefit to be capable of performing evaluation pointwise of the copula densities. We can see from Equation \ref{eq:density} that we are required to compute high dimensional derivatives of a composite function. In order to achieve this we utilise a specific multivariate chain rule result widely known as the Fa\`{a} di Bruno's Formula, see \cite{faa1857note} and discussions in for example \cite{constantine1996multivariate} and \cite{roman1980formula}.

\subsubsection{One-parameter Archimedean Members:}
\label{section:oneparam}
In this section we describe three of the one parameter multivariate Archimedean family copula models which have become popular model choices and are widely used for estimation. This is primarily due to there directly interpretable features. We select these three component members, the Clayton, Frank and Gumbel models, for our mixture models since they each contain differing tail dependence characteristics. Clayton provides lower tail dependence whilst Gumbel provides upper tail dependence. The Frank copula also provides dependence in the unit cube with elliptical contours with semi-major axis oriented at either $\pi/4$ or $3\pi/4$ depending on the sign of the copula parameter in the estimation. Therefore the Frank model component will allow us to capture parsimoniously potential negative dependence relationships between the currencies in the basket under study. Formulas for these copulae, as well as their respective generators, inverse generators and the d-th derivatives of their generators (required for the density evaluation) are given in Table~\ref{generators}. The explicit formulas for the d-th derivatives for all of the copulae in Table~\ref{generators} [Appendix B] were derived in \cite{hofert2012}.

\subsubsection{Two-parameter Archimedean Members via Outer Power Transforms}

In this section we also consider a more flexible generalization of the single parameter Clayton copula discussed above. To achieve this generalization we consider the outer-power transform of the Clayton copula, as discussed below which is based on a result in \cite{feller1971}.

\vspace{3mm}

\begin{defi} Outer power copula\\
The copula family generated by $\tilde \psi (t) = \psi(t^{\frac{1}{\beta}})$ is called an outer power family, where $\beta \in [1, \infty)$ and $\psi \in \Psi_\infty$ (the class of completely monotone Archimedean generators). 
\end{defi}

The proof of this follows from \cite{feller1971}, i.e. the composition of a completely monotone function with a non-negative function that has a completely monotone derivative is again completely monotone. Such copula model transforms were also studied in \cite{nelsen1997}, where they are referred to as a beta family associated with the inverse generator $\psi^{-1}$.

\begin{remark} The generator derivatives for the outer power transforms can be calculated using the base generator derivatives and Fa\`{a} di Bruno's multi-dimensional extension to the chain rule. The densities for the outer power Clayton copula in Table \ref{generators} [Appendix B] can thus be calculated using equation \ref{eq:density}.
\end{remark}

%% file: FittingtheModels_MA_14_Nov_2013_v3_0_0.tex
\section{Likelihood Based Estimation of the Mixture Copula Models}
\label{section:likelihood}

We begin this section with a discussion on the choices we make for the marginal distributions for each of the currencies specified in the baskets constructed for the high interest rate differentials and also the baskets for the low interest rate differentials.

In modelling parametrically the marginal features of the log return forward exchange rates, we wanted flexibility to capture a broad range of skew-kurtosis relationships as well as potential for sub-exponential heavy tailed features. In addition, we wished to keep the models to a selection which is efficient to perform inference and easily interpretable. We consider a flexible three parameter model for the marginal distributions given by the Log-Generalized-Gamma distribution (l.g.g.d.), see details in \cite{lawless1980inference} and \cite{consul1971log}.


%

The l.g.g.d. is a parametric model based on the generalized gamma distribution which is highly utilized in lifetime modelling and survival analysis. 
The log transformed g.g.d. random variable $Y = \ln X$ is given by the density 
\begin{equation} \label{EqnLGGD}
f_{Y}(y; k,u,b) = \frac{1}{b \Gamma(k)}\exp\left[k\left(\frac{y - u}{b} \right) - \exp\left(\frac{y-u}{b}\right) \right]
\end{equation}
with $u = \log(\alpha)$, $b = \beta^{-1}$ and the support of the l.g.g.d. distribution is $y \in \mathbb{R}$.

This flexible three parameter model is particularly interesting in the context of the marginal modelling we are considering since the LogNormal model is nested within the g.g.d. family as a limiting case (as $k \rightarrow \infty$). In addition the g.g.d. also includes the exponential model $(\beta=k=1)$, the Weibull distribution $(k=1)$ and the Gamma distribution $(\beta=1)$. Next we discuss how one can perform inference for the multivariate currency basket models using these marginal models and the mixture copula discussed previously.

\subsection{Two Stage Estimation: Inference Function For the Margins}

The inference function for margins (IFM) technique introduced in \cite{Joe1996} provides a computationally faster method for estimating parameters than Full Maximum Likelihood, i.e. simultaneously maximising all model parameters and produces in many cases a more stable likelihood estimation procedure. 
An alternative approach to copula model parameter estimation that is popular in the literature is known as the Maximum Partial Likelihood Estimator (MPLE) detailed in \cite{genest1995semiparametric}.

The procedure we adopt for likelihood based estimation is the two stage estimation known as Inference function for the Margins which is studied with regard to the asymptotic relative efficiency of the two-stage estimation
procedure compared with maximum likelihood estimation in \cite{Joe2005} and in \cite{Hafner2010}. It can be shown that the IFM estimator is consistent under weak regularity conditions. However, it is not fully efficient for the copula parameters. Nevertheless, it is widely used for its ease of implementation and efficiency in large data settings such as the models we consider in this study.

To complete this discussion on general IFM, before providing the MLE estimation expressions, we first note that in our study we fit copula models to the high interest rate basket and the low interest rate basket updated for each day in the period 02/01/1989 to 31/05/2013 using log return forward exchange rates at one month maturities for data covering both the previous 6 months and previous year as a sliding window analysis on each trading day in this period. Next we discuss briefly the marginal MLE estimations for the LogNormal and the l.g.g.d. models.

\subsubsection{Stage 1: Fitting the Marginal Distributions via MLE}

In the first step we fit the marginal distributions to the l.g.g.d. model.
The estimation for the three model parameters can be challenging due to the fact that a wide range of model parameters, especially for $k$, can produce similar resulting density shapes, see discussions in \cite{lawless1980inference}. To overcome this complication and to make the estimation efficient it is proposed to utilise a combination of profile likelihood methods over a grid of values for $k$ and perform profile likelihood based MLE estimation for each value of $k$, then for the other two parameters $b$ and $u$. The differentiation of the profile likelihood for a given value of $k$ produces the system of two equations
\begin{equation}
\exp(\tilde\mu) = \left[ \frac{1}{n}\sum_{i=1}^n \exp\left(\frac{y_i}{\tilde\sigma\sqrt{k}}\right) \right]^{\tilde\sigma \sqrt{k}}\\
\hspace{10mm} ; \hspace{10mm}
\frac{\sum_{i=1}^n y_i \exp\left(\frac{y_i}{\tilde\sigma\sqrt{k}}\right)}{\sum_{i=1}^n \exp\left(\frac{y_i}{\tilde\sigma\sqrt{k}}\right)} - \overline{y} - \frac{\tilde\sigma}{\sqrt{k}} = 0
\label{lggd_mle}
\end{equation}
with $n$ the number of observations, $y_i = \log x_i$ and the parameter transformations $\tilde\sigma = \frac{b}{\sqrt{k}}$ and $\tilde\mu = u + b\ln k$. The second equation is solved directly via a simple root search for the estimation of $\tilde{\sigma}$ and then substitution into the first equation provides the estimation of $\tilde{\mu}$. Note, for each value of $k$ we select in the grid, we get the pair of parameter estimates $\tilde\mu$ and $\tilde\sigma$, which can then be plugged back into the profile likelihood to make it purely a function of $k$, with the estimator for $k$ then selected as the one with the maximum likelihood score.

\subsubsection{Stage 2: Fitting the Mixture Copula via MLE}

In order to fit the copula model the parameters are estimated using maximum likelihood on the data after conditioning on the selected marginal distribution models and their corresponding estimated parameters obtained in Stage 1. These models are utilised to transform the data using the cdf function with the l.g.g.d. MLE parameters ($\hat k$, $\hat u$ and $\hat b$).

Therefore, in this second stage of MLE estimation we aim to estimate either the one parameter mixture of C-F-G components with parameters ${\bf\underline\theta} = (\rho_{clayton}, \rho_{frank}, \rho_{gumbel}, \lambda_{clayton}, \lambda_{frank}, \lambda_{gumbel})$, the one parameter mixture of C-G components with parameters ${\bf\underline\theta} = (\rho_{clayton}, \rho_{gumbel}, \lambda_{clayton}, \lambda_{gumbel})$  or the two parameter outer power transformed OpC with parameters ${\bf\underline\theta} = (\rho_{clayton}, \beta_{clayton})$. This is achieved in each case by the conditional maximum likelihood. To achieve this we need to maximise the log likelihood expressions for the mixture copula models, which in our framework are given generically by the following function for which we need to find the mode,
\begin{equation}
l({\bf\underline\theta}) = \sum_{i=1}^n log \; c(F_1(X_{i1};\hat\mu_1, \hat\sigma_1), \dots, F_d(X_{id};\hat\mu_d, \hat\sigma_d))  \; + \; \sum_{i=1}^n \sum_{j=1}^d log  \;f_j(X_{ij};\hat\mu_j, \hat\sigma_j)
\label{loglik}
\end{equation}
with respect to the parameter vector ${\bf\underline\theta}$.

For example in the case of the Clayton-Frank-Gumbel mixture copula we need to maximise on the log-scale the following expression.
\begin{equation}
\begin{aligned}
l({\bf\underline\theta}) = \sum_{i=1}^n log \; \big\lbrack & \lambda_C * \left(c_{\rho_C}^C \left(F_1 \left(X_{i1};\hat\mu_1, \hat\sigma_1 \right) \dots, F_d \left(X_{id};\hat\mu_d, \hat\sigma_d \right)\right) \right)  \\ 
&+ \lambda_F* \left(c_{\rho_F}^F \left(F_1 \left(X_{i1};\hat\mu_1, \hat\sigma_1 \right) \dots, F_d \left(X_{id};\hat\mu_d, \hat\sigma_d \right)\right) \right)  \\ 
&+ \lambda_G* \left(c_{\rho_G}^G \left(F_1 \left(X_{i1};\hat\mu_1, \hat\sigma_1 \right) \dots, F_d \left(X_{id};\hat\mu_d, \hat\sigma_d \right)\right) \right) \big\rbrack
\end{aligned}
\label{cgloglik}
\end{equation}
This optimization is achieved via a gradient descent iterative algorithm which was found to be quite robust given the likelihood surfaces considered in these models with the real data. To illustrate this point, at this stage it is instructive to present some examples of the shapes of the profile likelihoods that are being optimized over for some of the important copula model parameters in the C-F-G mixture example for a 6 month window of data randomly selected from the data set for both the high interest rate basket and the low interest rate basket. 


\begin{remark}
A number of randomly selected copula fits from across the time period were analysed. The profile likelihood plots showed level sets which were not particularly complicated or multi-modal. Therefore the simple maximum likelihood estimation approach is sufficient here. If the likelihood hyper-surface was more complicated one could use the expectation-maximisation algorithm.
\end{remark}



\subsection{Goodness-of-Fit Tests}
\label{section:AIC}
In this section we briefly comment on the model selection aspects of the analysis we undertook. We first undertook a process of fitting the marginal LogNormal model to all of the 33 currencies considered in the analysis over a sliding window of 6 months and 1 year. For each of these fits we then performed a formal hypothesis test via Kolmogorov-Smirnov, at a level of significance of 5\%, to test the validity of the LogNormal assumption.
We found strong evidence to reject the null systematically for a few important developing countries' marginal models. Since LogNormal is a subfamily of l.g.g.d. we proceeded to use the l.g.g.d. marginal distribution family.
The resulting estimated model fits were significantly improved under the l.g.g.d. models.



In terms of the selection of the copula mixture models, between the three component mixture C-F-G model versus the two component mixture C-G model versus the two parameter OpC model, we used a scoring via the AIC.
One could also use the Copula-Information-Criterion (CIC), see \cite{Gronneberg2010} for details. The results are presented for this comparison in Figure~\ref{AIC_diff_49} [Appendix A], which shows the differentials between AIC for C-F-G versus C-G and C-F-G versus Op-C for each of the currency baskets.
We can see it is not unreasonable to consider the C-F-G model for this analysis, since the mean difference between the AIC scores for the C-F-G and the C-G models for the high interest rate basket is 2.33 and for the low interest rate basket is 2.25 in favour of the C-F-G. The C-F-G copula model also provides a much better fit when compared to the OpC model, as shown by the mean difference between the AIC scores of 14.43 for the high interest rate basket and 13.36 for the low interest rate basket.

\FloatBarrier

%% file: Joint_Tail_Risk_Exposure_MA_28_Nov_2013_v4_0_0.tex
\section{Joint Tail Exposure in the Carry Trade via Extremal Tail Dependence Coefficient}
\label{joint_tail_risk_exposure}



In order to fully understand the tail risks of joint exchange rate movements present when one invests in a carry trade strategy we can look at both the downside extremal tail exposure and the upside extremal tail exposure within the long and the short baskets that comprise the strategy. The downside tail exposure can be seen as the crash risk of the basket, i.e. the risk that one will suffer large joint losses from each of the currencies in the basket. These losses would be the result of joint appreciations of the currencies one is short in the low interest rate basket and/or joint depreciations of the currencies one is long in the high interest rate basket. The downside tail exposures are thus characterised by the conditional probabilities that one or more currencies in the long/short basket depreciates/appreciates beyond an extreme threshold given that the remaining currencies in the long/short basket depreciate/appreciate beyond this threshold.

The upside tail exposure is the risk that one will earn large joint profits from each of the currencies in the basket. These profits would be the result of joint depreciations of the currencies one is short in the low interest rate basket and/or joint appreciations of the currencies one is long in the high interest rate basket. The upside tail exposures are thus characterised by the conditional probabilities that one or more currencies in the short/long basket depreciates/appreciates beyond an extreme threshold given that the remaining currencies in the short/long basket depreciate/appreciate beyond this threshold.

We can formalise this notion of the dependence behaviour in the extremes of the multivariate distribution through the concept of tail dependence. The tail dependence coefficient is defined as the conditional probability that a random vector exceeds a certain threshold (in the limit as the threshold goes to infinity) given that the remaining components of the random vector in the joint distribution has exceeded this threshold. This concept had previously been considered in the bivariate setting and was recently extended to the multivariate setting by \cite{de2012multivariate}. Now one may accurately interpret the tail dependence present between sub-vector partitions of the multivariate random vector of currency exchange rates in either basket with regard to joint tail dependence behaviours. In the context of the applications we consider in this paper, this allows us to examine the probability that any subvector of the log return forward exchange rates for the basket of currencies will exceed a certain threshold given that the log return forward exchange rates for the remaining currencies in the basket have exceeded this threshold, in particular thresholds that are placing an interest in the tails of the multivariate distribution. The interpretation of such results is then directly relevant to assessing the chance of large adverse movements in multiple currencies which could potentially increase the risk associated with currency carry trade strategies significantly, compared to risk measures which only consider the marginal behaviour in each individual currency. We thus consider these tail upside and downside exposures as features that can show that even though average profits may be made from the violation of UIP, it comes at significant tail exposure.

Below we give the definitions of upper and lower tail dependence, as well as the explicit generalised multivariate expressions for Archimedean copulae, equations~\ref{eq:archmuppertd} and ~\ref{eq:archmlowertd}, derived in \cite{de2012multivariate}.

\vspace{3mm}

\begin{defi} Generalized Upper Tail Dependence Coefficient \\
Let $X = (X_1,..., X_d)^T$ be a d dimensional random vector
with marginal distribution functions $F_1, ..., F_d$. The coefficient of upper tail dependence is defined as:
\begin{equation}
\begin{aligned}
\lambda_u^{1,...,h|h+1,...,d} &= \lim_{\nu \rightarrow 1-} P\left( X_1 > F^{-1}(\nu),...,X_h > F^{-1}(\nu) | X_{h+1} > F^{-1}(\nu), ..., X_d > F^{-1}(\nu) \right) \\
&= \lim_{t \rightarrow 0^+} \frac{\sum_{i=1}^d \left( \binom{d}{d-i} i (-1)^{i} \left[ \psi^{-1'}  (it) \right] \right) }{\sum_{i=1}^{d-h} \left( \binom{d-h}{d-h-i} i (-1)^i \left[ \psi^{-1'} (it) \right] \right)}
\end{aligned}
\label{eq:archmuppertd}
\end{equation}
\end{defi}

\vspace{3mm}

\begin{defi}Generalized Lower Tail Dependence Coefficient \\
Let $X = (X_1,..., X_d)^T$ be a d dimensional random vector
with marginal distribution functions $F_1, ..., F_d$. The coefficient of lower tail dependence is defined as:
\begin{equation}
\begin{aligned}
\lambda_l^{1,...,h|h+1,...,d} &= \lim_{\nu \rightarrow 0+} P \left( X_1 < F^{-1}(\nu),...,X_h < F^{-1}(\nu) | X_{h+1} < F^{-1}(\nu), ..., X_d < F^{-1}(\nu) \right) \\
&= \lim_{t \rightarrow \infty} \frac{d}{d-h} \frac{\psi^{-1'} (dt)}{\psi^{-1'} ((d-h)t)}
\end{aligned}
\label{eq:archmlowertd}
\end{equation}
\end{defi}

\begin{remark}
The downside exposures are thus the upper tail dependence in the high interest rate basket and the lower tail dependence in the low interest rate basket. The upside exposures are the lower tail dependence in the high interest rate basket and the upper tail dependence in the low interest rate basket.
\end{remark}


%

%% file: Results_MA_14_Nov_2013_v3_0_0.tex
\section{Results}

In this section we present a detailed analysis of the estimation of the marginal distributional models and the mixture copula models for both the high interest rate basket and the low interest rate basket. Firstly, we investigate the properties of the marginal distributions of the exchange rate log-returns for the 33 currencies. We then interpret the time-varying dependence characteristics of the fitted copula models to the high interest rate basket and the low interest rate basket across the period 02/01/1989 to 31/05/2013.

\FloatBarrier



\subsection{Estimation Results for the Copula Modelling}

We utilised each of the l.g.g.d. marginal distribution fits for a given day's set of currencies in the high interest rate and low interest rate baskets to analyse the joint multivariate features. To achieve this for each of the currencies, the exchange rate log-return data was transformed via the l.g.g.d. marginal model's distribution function. Thus producing pseudo data which is approximately marginally uniform [0, 1].  Then the mixture Clayton-Frank-Gumbel copula (denoted C-F-G ), the mixture Clayton-Gumbel copula (denoted C-G) and the outer-power Clayton copula were fitted each day to a sliding window of 6 months and one year log-returns data for both the high interest rate and low interest rate baskets. Below we will examine the time-varying parameters of the maximum likelihood fits of this mixture C-F-G copula model, since as was shown in Section \ref{section:AIC} the C-F-G copula provided the best fit on average.

In this analysis there are several attributes to be considered for the mixture copula model, such as the relevant copula structures for the high and low interest rate baskets and how these copula dependence structures may change over time. In addition, there is the strength of the tail dependence in each currency basket and how this changes over time, especially in periods of heightened market volatility. The first of these attributes we will consider to be a structure analysis studying the relevant forms of dependence in the currency baskets and the second of these attributes that we shall study will be the strength of dependence present in the currency baskets, given the particular copula structures in the mixture, which is considered as tail upside/downside exposure of a carry trade over time.

Therefore we first consider the structural components of the multivariate copula model. To achieve this, we begin with a form of model selection in a mixture context, in which we consider the estimated relative contributions of each of the copula components (and their associated dependence features) to the joint relationship in the high and low interest rate currency baskets over time. This is reflected in the estimated mixture component weights, which can be seen in Figures~\ref{lambda_high_49} and~\ref{lambda_low_49} [Appendix A] for the high interest rate basket and low interest rate basket respectively. The $\lambda$ values show the relevance of each of the component copulae to the data. Thus a small $\lambda$ value indicates the lack of a need for that particular copula component in order to model the associated 6 months or one year block of data. In contrast, a $\lambda$ value for the Gumbel component very close to 1 indicates the block of data could be well modelled by a Gumbel copula alone. Hence, these plots convey the time varying significance of hypotheses about the presence of upper and lower tail dependence in each of the baskets over time. Examining these plots shows that in general the Clayton mixture weight tends to be lower when the Gumbel mixture weight is higher. We can also see that the Frank copula is systematically present in the mixture. In addition, we see that in the periods of high market volatility we observed differences in the relevant upper and lower tail dependence structural attributes when comparing the high versus low interest rate baskets. That is, there is an asymmetric tendency for the presence of particular copula components over time when comparing the high and low interest rate baskets. The implications of this will be discussed in further detail in Section~\ref{discussion}.



In terms of the second attribute, the strength of the copula dependence, we analyse this in several ways. Firstly through an analysis of the estimated copula parameter components over time, then through an analysis of the transformation of these copula parameters to rank correlations and finally through an analysis of the multivariate strength of the mixture copula tail dependence over time.

The individual component copula parameters can be seen in Figures~\ref{rho_high_49} and~\ref{rho_low_49} [Appendix A] for the high interest rate basket and low interest rate basket respectively. The strength of the copula parameters in the baskets shows a large degree of variance during the period 02/01/1989 to 31/05/2013.




The measure of concordance as captured by Kendall's tau is decomposed in this analysis according to each of the mixture components, scaled by the mixture weights $\lambda$, and can be seen in Figure~\ref{tau_high_49} for the high interest rate basket and Figure~\ref{tau_low_49} [Appendix A] for the low interest rate basket. These plots provide a more intuitive picture of the time-varying contributions of the individual copulae to the dependence structure present in each of the baskets. Interestingly, we see the rank correlation contribution from the Frank copula indicates the presence of negative as well as positive rank correlations. In addition, as discussed with the mixture weights, there is perhaps some asymmetry present between the high and low interest rate baskets over time.

\begin{figure}[p]
\centering
\includegraphics[width =\textwidth , height=90mm]{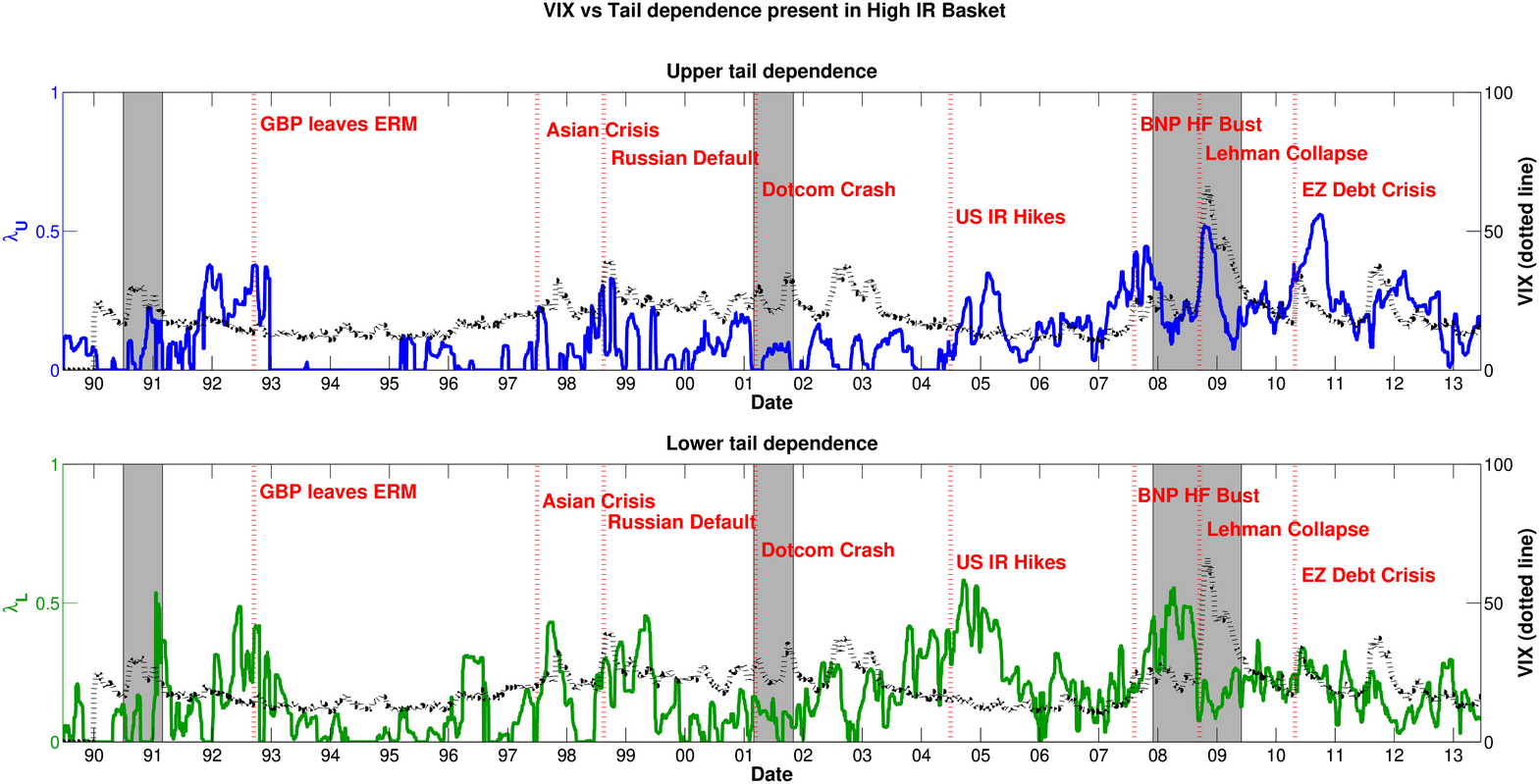}
\caption{Comparison of Volatility Index (VIX) with upper and lower tail dependence of the high interest rate basket. US NBER recession periods are represented by the shaded grey zones. Some key crisis dates across the time period are labelled.}
\label{VIX_vs_TD_high_49}
\end{figure}

\begin{figure}[p]
\centering
\includegraphics[width =\textwidth , height=90mm]{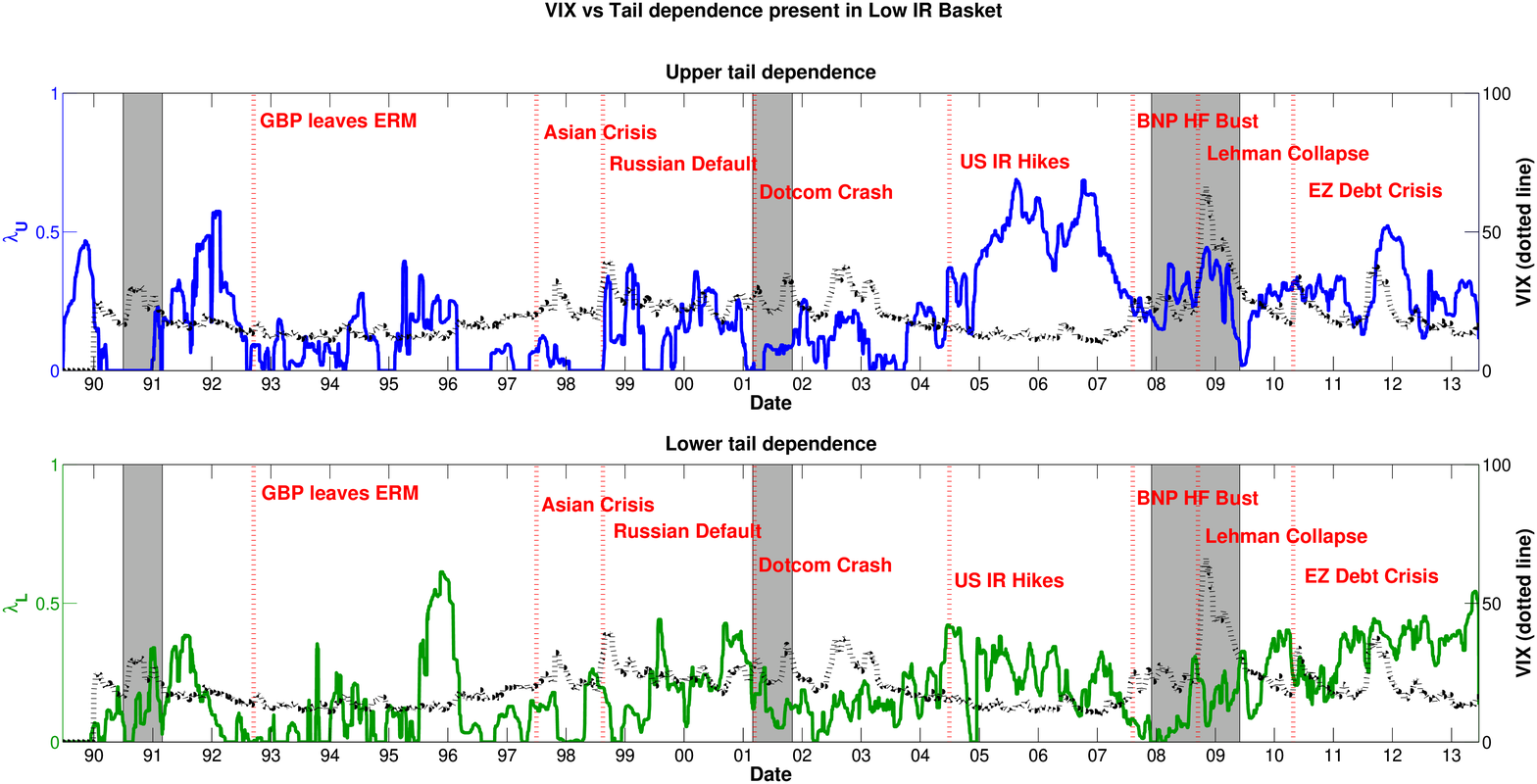}
\caption{Comparison of Volatility Index (VIX) with upper and lower tail dependence of the low interest rate basket. US NBER recession periods are represented by the shaded grey zones. Some key crisis dates across the time period are labelled.}
\label{VIX_vs_TD_low_49}
\end{figure}

Perhaps the most interesting and revealing representation of the tail dependence characteristics of the currency baskets can be seen in Figures~\ref{VIX_vs_TD_high_49} and \ref{VIX_vs_TD_low_49}. Here we can see that there are indeed periods of heightened upper and lower tail dependence in the high interest rate and the low interest rate baskets. There is a noticeable increase in upper tail dependence in the high interest rate basket at times of global market volatility. Specifically, during late 2007, i.e. the global financial crisis, there is a sharp peak in upper tail dependence. Preceding this, there is an extended period of heightened lower tail dependence from 2004 to 2007, which could tie in with the building of the leveraged carry trade portfolio positions. This period of carry trade construction is also very noticeable in the low interest rate basket through the very high levels of upper tail dependence.

In understanding this analysis we note that Figures~\ref{VIX_vs_TD_high_49} and~\ref{VIX_vs_TD_low_49} show the probability that one currency in the basket will have a move above/below a certain extreme threshold given that the other currencies have had a move beyond this threshold.

%

To illustrate the relationship between heightened periods of significant upper and lower tail dependence features over time and to motivate the clear asymmetry present in the upper and lower tail dependence features between the high and low interest rate baskets over time we consider a further analysis. In particular we compare in Figures~\ref{VIX_vs_TD_high_49} and~\ref{VIX_vs_TD_low_49} the tail dependence plotted against the VIX volatility index for the high interest rate basket and the low interest rate basket respectively for the period under investigation. The VIX is a popular measure of the implied volatility of S\&P 500 index options - often referred to as the \emph{fear index}. As such it is one measure of the market's expectations of stock market volatility over the next 30 days. We can clearly see here that in the high interest rate basket there are upper tail dependence peaks at times when there is an elevated VIX index, particularly post-crisis. However, we would not expect the two to match exactly since the VIX is not a direct measure of global FX volatility. We can thus conclude that investors' risk aversion clearly plays an important role in the tail behaviour of high interest rate currencies and more importantly in their dependence structure. This statement can also be associated to the globalization of financial markets and the resulting increase of the contagion risk between countries. This conclusion corroborates some of the recent literature results with regards to the skewness and the kurtosis features characterizing the currency carry trade portfolios \citet{Farhi2008,Brunnermeier2008, Menkhoff2012}.

\subsection{Understanding the Tail Exposure associated with the Carry Trade and its Role in the UIP Puzzle}

In the existing literature, the risks associated with the carry trade have been analysed from the marginal perspective or from an aggregated portfolio perspective, i.e. taking into account the mean, variance, skewness and kurtosis of portfolio returns, see \citet{Menkhoff2012,Lustig2011,Brunnermeier2008}. However, until now the tails in the dependence structure within each basket of the carry trade have been overlooked as an additional source of risk. It is important to note that in our paper we study the extremal tail exposure, not the intermediate tail risk (see \citet{hua2012strength,hua2012intermediate}), and hence we focus on the relative effect of the extreme exposures on portfolio returns. In order to have a better grasp of the risk globally associated to a currency carry trade portfolio, which is derived from the differential of interest rates, we propose to compare the relative tail exposure adjusted returns for the two components of such a portfolio, namely the high and the low interest rates baskets.

As was discussed in Section~\ref{joint_tail_risk_exposure}, the tail exposures associated with a currency carry trade strategy can be broken down into the upside and downside tail exposures within each of the long and short carry trade baskets.
The downside tail exposure can be seen as the crash risk of the basket, i.e. the risk that one will suffer large joint losses from each of the currencies in the basket. These losses would be the result of joint appreciations of the currencies one is short in the low interest rate basket and/or joint depreciations of the currencies one is long in the high interest rate basket. The downside tail exposures are thus characterised by the conditional probabilities that one or more currencies in the long/short basket depreciates/appreciates beyond an extreme threshold given that the remaining currencies in the long/short basket depreciate/appreciate beyond this threshold.
The upside tail exposure is the risk that one will earn large joint profits from each of the currencies in the basket. These profits would be the result of joint depreciations of the currencies one is short in the low interest rate basket and/or joint appreciations of the currencies one is long in the high interest rate basket. The upside tail exposures are thus characterised by the conditional probabilities that one or more currencies in the short/long basket depreciates/appreciates beyond an extreme threshold given that the remaining currencies in the short/long basket depreciate/appreciate beyond this threshold.

The downside relative exposure adjusted returns are obtained by multiplying the monthly portfolio returns by one minus the upper and the lower tail dependence present respectively in the high interest rate basket and the low interest rate basket at the corresponding dates. The upside relative exposure adjusted returns are obtained by multiplying the monthly portfolio returns by one plus the lower and upper tail dependence present respectively in the high interest rate basket and the low interest rate basket at the corresponding dates. Note that we refer to these as relative exposure adjustments only for the tail exposures since we do not quantify a market price per unit of tail risk. However, this is still informative as it shows a decomposition of the relative exposures from the long and short baskets with regard to extreme events.

\begin{figure}
\centering
\includegraphics[width =\textwidth , height=90mm]{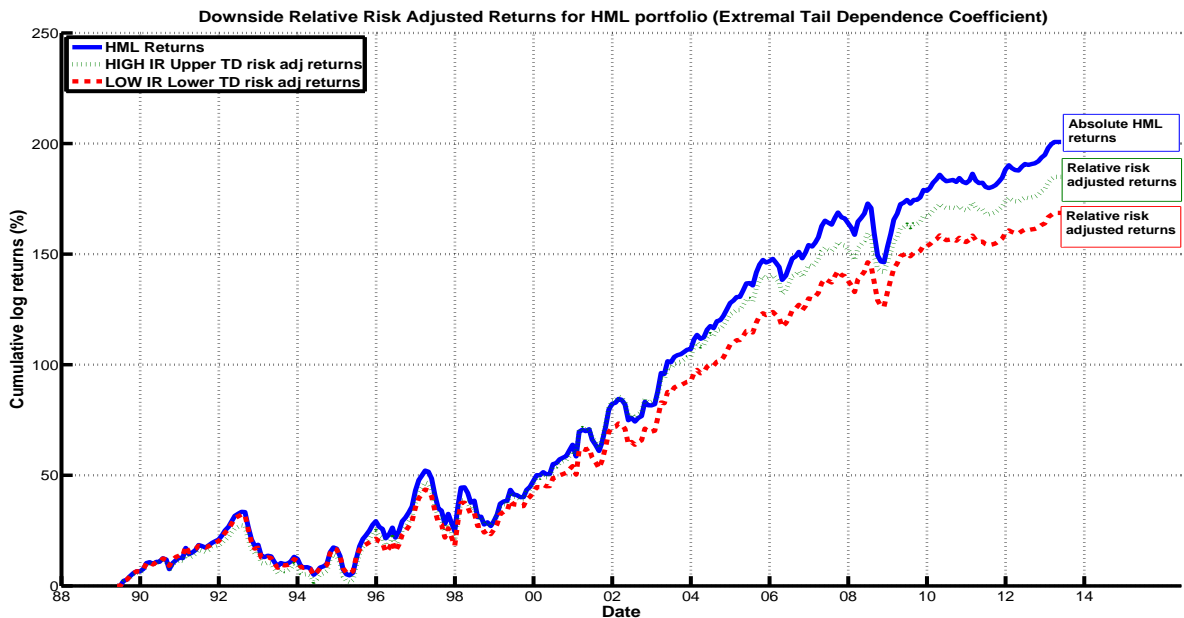}
\caption{Cumulative log returns of the carry trade portfolio (HML = High interest rate basket Minus Low interest rate basket). Downside exposure adjusted cumulative log returns using upper/lower tail dependence in the high/low interest rate basket are shown for comparison.}
\label{risk_adj_Downside}
\end{figure}

\begin{figure}
\centering
\includegraphics[width =\textwidth , height=90mm]{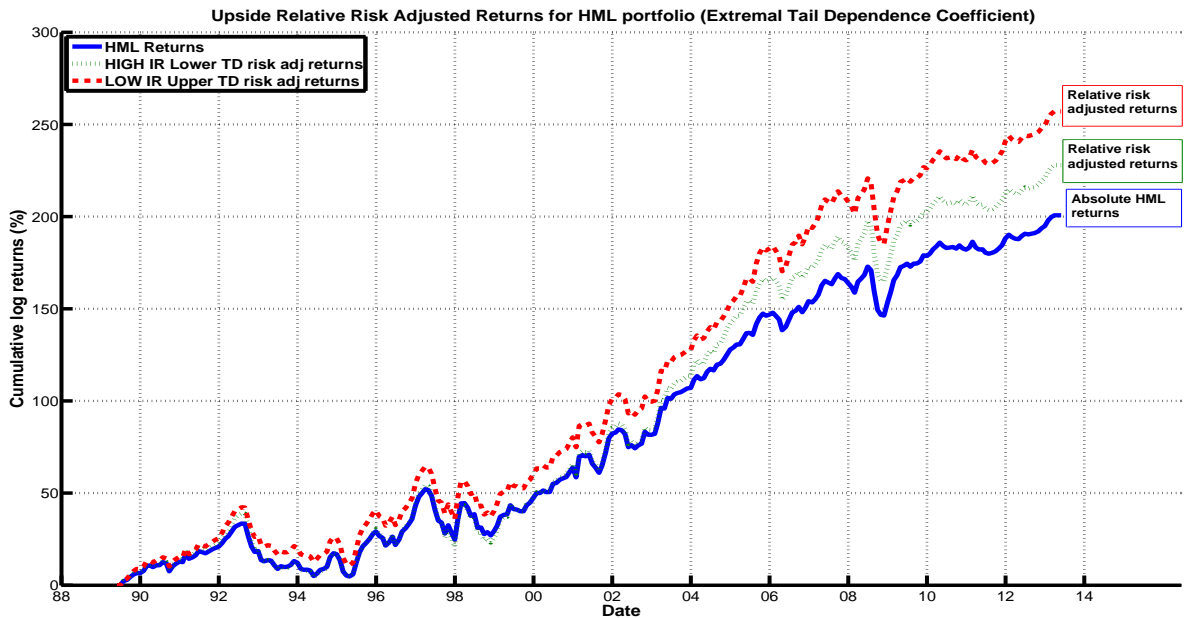}
\caption{Cumulative log returns of the carry trade portfolio (HML = High interest rate basket Minus Low interest rate basket). Upside exposure adjusted cumulative log returns using lower/upper tail dependence in the high/low interest rate basket are shown for comparison.}
\label{risk_adj_Upside}
\end{figure}

As can been seen in Figure~\ref{risk_adj_Downside}, the relative adjustment to the absolute cumulative returns for each type of downside exposure is greatest for the low interest rate basket. This is interesting because intuitively one would expect the high interest rate basket to be the largest source of tail exposure. However, one should be careful when interpreting this plot, since we are looking at the extremal tail exposure. The analysis would change if one considered the intermediate tail risk exposure, where the marginal effects become significant.
Similarly, Figure~\ref{risk_adj_Upside} shows the relative adjustment to the absolute cumulative returns for each type of upside exposure is greatest for the low interest rate basket. The same interpretation as for the downside relative exposure adjustments can be made here for upside relative exposure adjustments.

%% file: Discussion_MA_14_Nov_2013_v3_0_0.tex
\section{Discussion}
\label{discussion}

This article sheds light on an important topic that has been discussed in the recent financial literature which is the Uncovered interest rate parity (UIP) puzzle. This market phenomenon is particularly interesting from a theoretical standpoint as well as for the understanding of financial market mechanisms. It has been demonstrated that the currency markets were indeed not respecting empirically a fundamental relation in finance which connects the currency exchange rates and the interest rates associated with two different countries. The main contribution of this article has been to propose a rigorous statistical modelling approach which captures the specific statistical features of both the individual currency log-return distributions as well as the joint features such as the dependence structures prevailing between all the exchange rates. This has helped to understand a previously unexplored stochastic feature of the UIP violation.

In achieving this goal, we first assessed the marginal statistical features of each of the 33 currencies on an assumed locally stationary sliding window of six months, over all the trading days in the period 02/01/1989 to 31/05/2013. 
This showed that the flexible l.g.g.d. model was capable of capturing a range of skew-kurtosis relationships present in the real data.

Having modelled the marginal attributes of the high and low interest rate currency baskets over time adequately, it was then our main emphasis to assess the multivariate dependence features of the currency baskets. In particular how this may change over time within a given basket, where we were particularly interested in the effect of the composition of the basket over time, and the response of the multivariate dependence features of the modelled basket and how it may respond in periods of heightened market volatility versus more stable periods. In addition to this within basket temporal analysis, from the perspective of undertaking a currency carry trade strategy, we would need to consider the relative relationships between the temporal dependence features of the high interest rate and low interest rate currency baskets. 
We demonstrate several interesting features from our model fits relating to asymmetries between the high and low interest rate baskets over time, especially during periods of high volatility in global markets. One way we ascertained such periods was through a graphical comparison of the VIX versus features of the multivariate dependence relationships we modelled. Importantly we found substantial evidence to support arguments for time varying behaviours both in the structural dependence hypotheses posed about the currency baskets, as captured by the relevant contributing copula components to the multivariate mixture model. As well as substantial evidence for significant tail dependence features in both the high and low interest rate baskets, which again displayed interesting asymmetries between the high and low interest rate baskets over time. 

The financial interpretation of the significance of these findings is related to the fact that it demonstrates that historically average rewards from a currency carry trade portfolio can be exposed to a significant risk of large losses arising from joint adverse movements in the currencies that would typically comprise the high and low interest rate baskets that an investor would go long and short when performing a currency carry trade. Hence, we conclude that our second contribution to the literature has been to rigorously demonstrate that such assertions relating to the profitability of the currency carry trade are failing to appropriately take into consideration an important component of the risk which characterizes these types of portfolios of currencies named carry trade portfolios, i.e. the upper and lower tail dependence features that we show are empirically present consistently in historic data.

We conclude that indeed the copula theory we employed allows us to demonstrate statistically that beyond the intrinsic risk associated to high interest rate countries (which are generally paying higher interest rates to compensate for a higher risk) typically studied in the literature from a marginal perspective, another source of risk plays an important role. This second source of risk is related to the dependence structures linking these high interest rate currencies, more specifically the significant tail dependence features observed in our model analysis. We indeed proved through a mixture of Archimedean copulae the significant presence of tail dependence among high interest rate currencies which could have dramatic consequences on the carry trade portfolio's risk profile when accounted for appropriately in risk reward analysis. As a matter of fact, the tail dependence directly influences the diversity of the assets and thus reduces the appealing convergence property stated by the modern portfolio theory. 

Said differently, our copula based probabilistic modelling approach allows us to demonstrate that besides the intrinsic risk associated to each particular high interest rate currency, another factor constitutes a determining source of risk which turns out to be the level of risk aversion prevailing in the market. It was demonstrated in our analysis that both upper and lower tail dependence features displayed significant association and asymmetries with each other between the high and low interest rate baskets during periods of relative financial stability versus periods of heightened market volatility.

These tail dependence features in the high interest rate basket were significantly increasing during crisis periods leading to an increased exposure associated with utilising such currency baskets (which were no longer diversified due to the presence of significant tail dependence features) in a carry trade. That being said, a rational portfolio manager's natural risk aversion tells them that they should receive an additional remuneration in order to offset any additional sources of risk associated to an investment. Therefore, to properly assess the profitability of the currency carry trade, such tail dependence features should be incorporated into the analysis of the risk-reward. To conclude, our paper rigorously tempers the too often claimed attractiveness of the currency carry trade and provides to investors a risk management tool in order to control and monitor the risk contained in such positions.

%% file: appendixA_MA_14_Nov_2013_v3_0_0.tex
\section*{Appendix A}

\FloatBarrier

\begin{figure}[ht!]
\centering
\includegraphics[height=90mm, keepaspectratio]{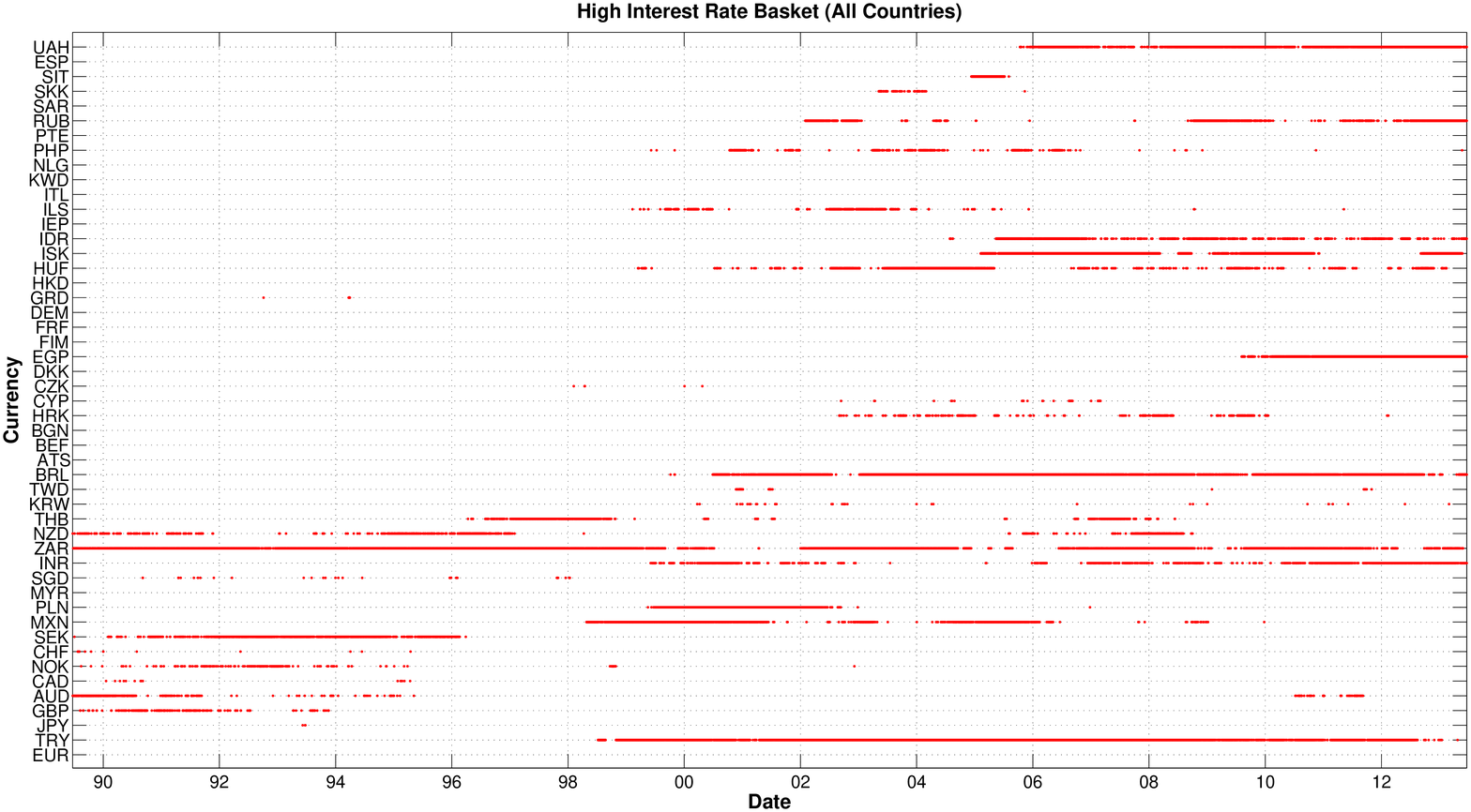}	
\caption{Highest interest rate basket composition over time. The basket is updated on a daily basis to contain the highest n/5 currencies sorted by interest rate, where n is the number of currencies available in the dataset on that day.}
\label{b5_49}
\end{figure}
\begin{figure}[ht!]
\centering
\includegraphics[height=90mm, keepaspectratio]{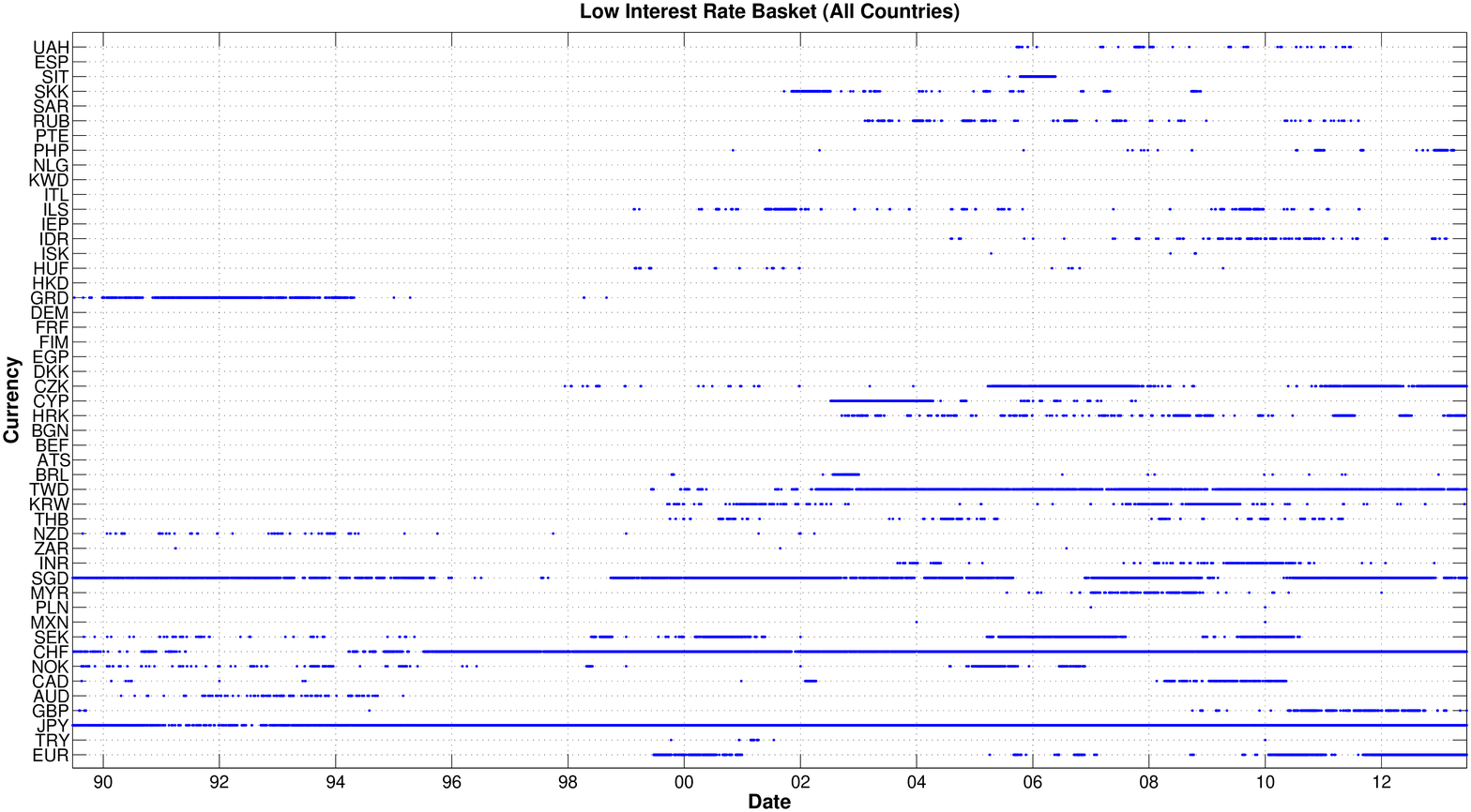} 
\caption{Lowest interest rate basket composition over time. The basket is updated on a daily basis to contain the lowest n/5 currencies sorted by  interest rate, where n is the number of currencies available in the dataset on that day.}
\label{b1_49}
\end{figure}

\begin{figure}
\centering
\includegraphics[width =\textwidth , height=90mm]{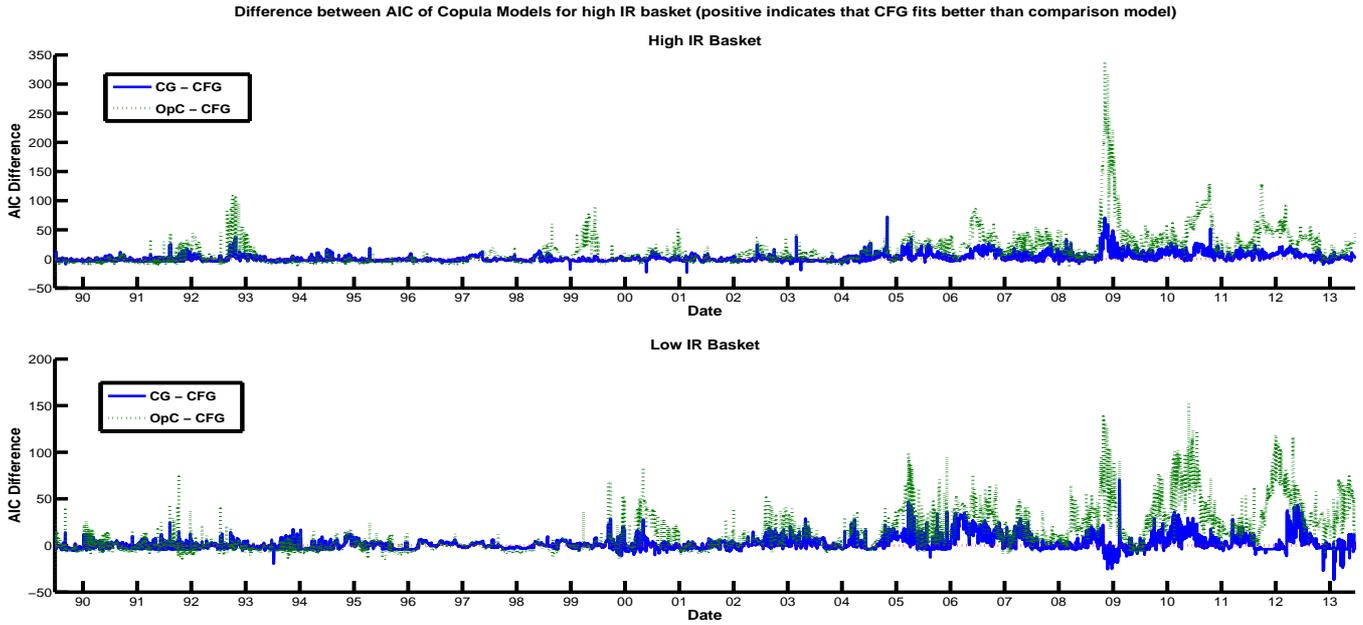}
\caption{Differences between AIC for C-F-G vs C-G and C-F-G vs Op-C for 6 month blocks on high and low IR basket. A positive value indicates that the CFG model is a better fit than the comparison model.}
\label{AIC_diff_49}
\end{figure}

\begin{figure}[p]
\centering
\includegraphics[width =\textwidth , height=90mm]{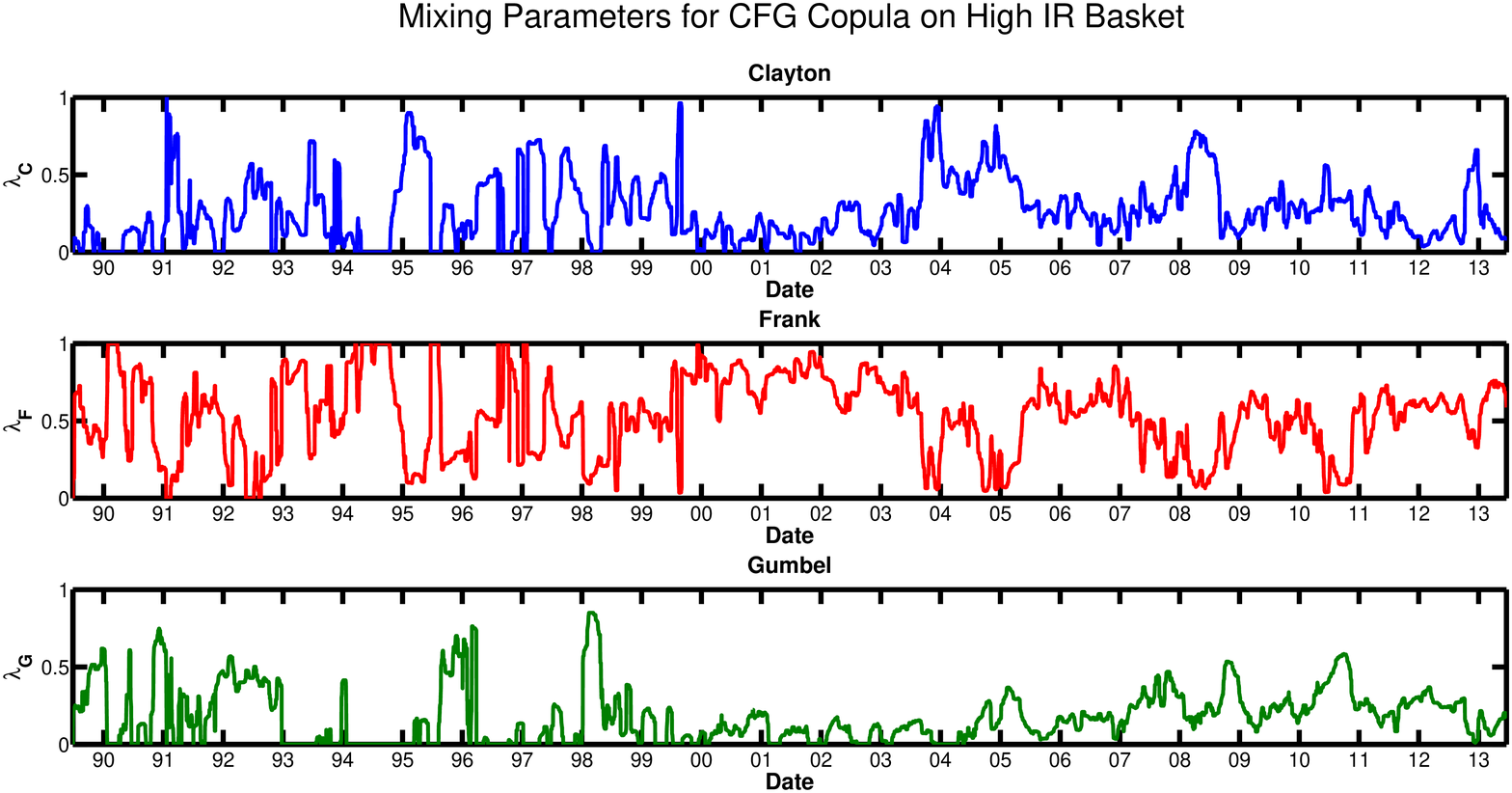}			
\caption{$\lambda$ Mixing proportions of the respective Clayton, Frank and Gumbel copulae
on the high interest rate basket.}
\label{lambda_high_49}
\end{figure}

\begin{figure}[p]
\centering
\includegraphics[width =\textwidth , height=90mm]{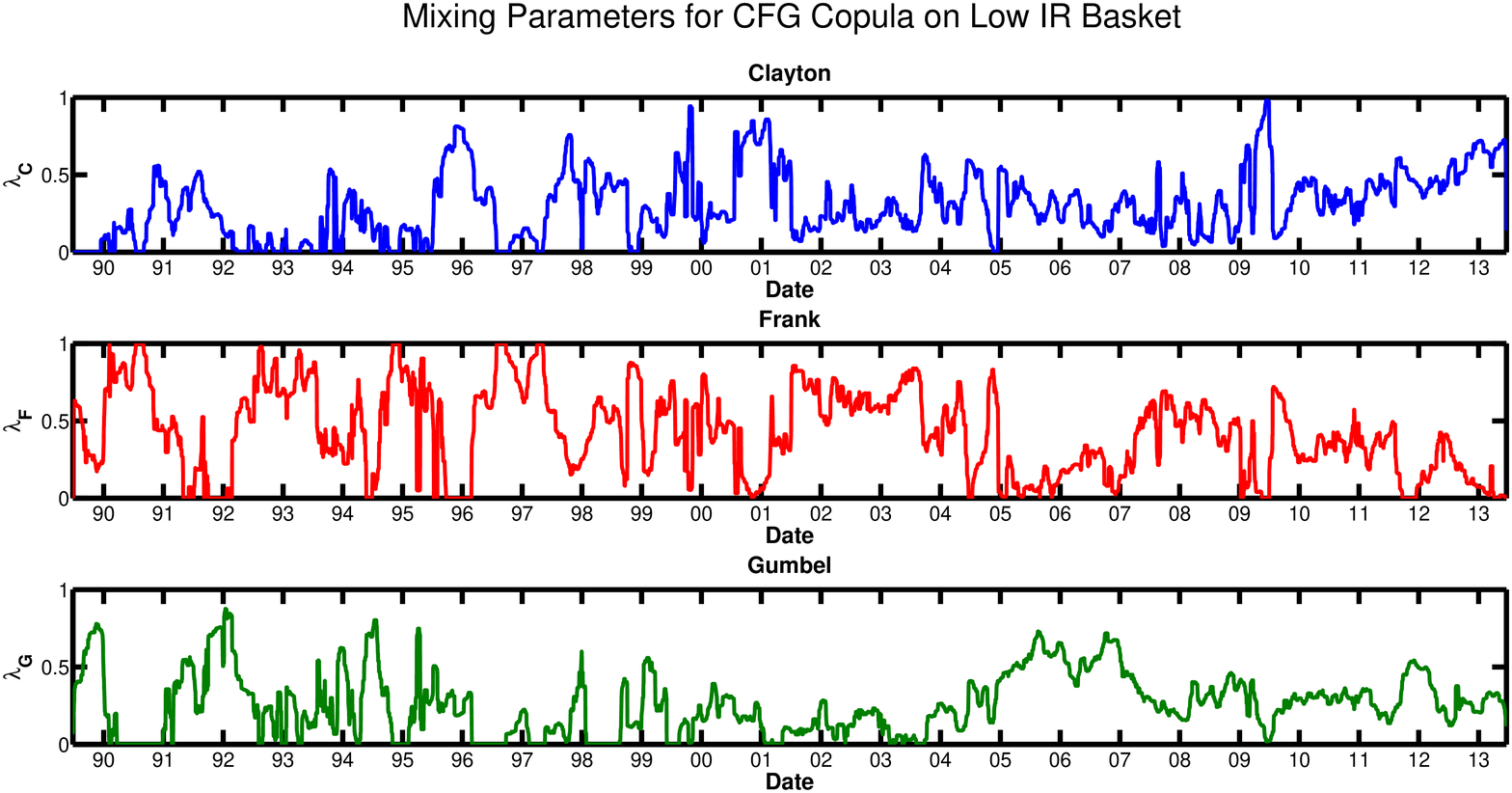}
\caption{$\lambda$ Mixing proportions of the respective Clayton, Frank and Gumbel copulae
on the low interest rate basket.}
\label{lambda_low_49}
\end{figure}

\begin{figure}[p]
\centering
\includegraphics[width =\textwidth , height=90mm]{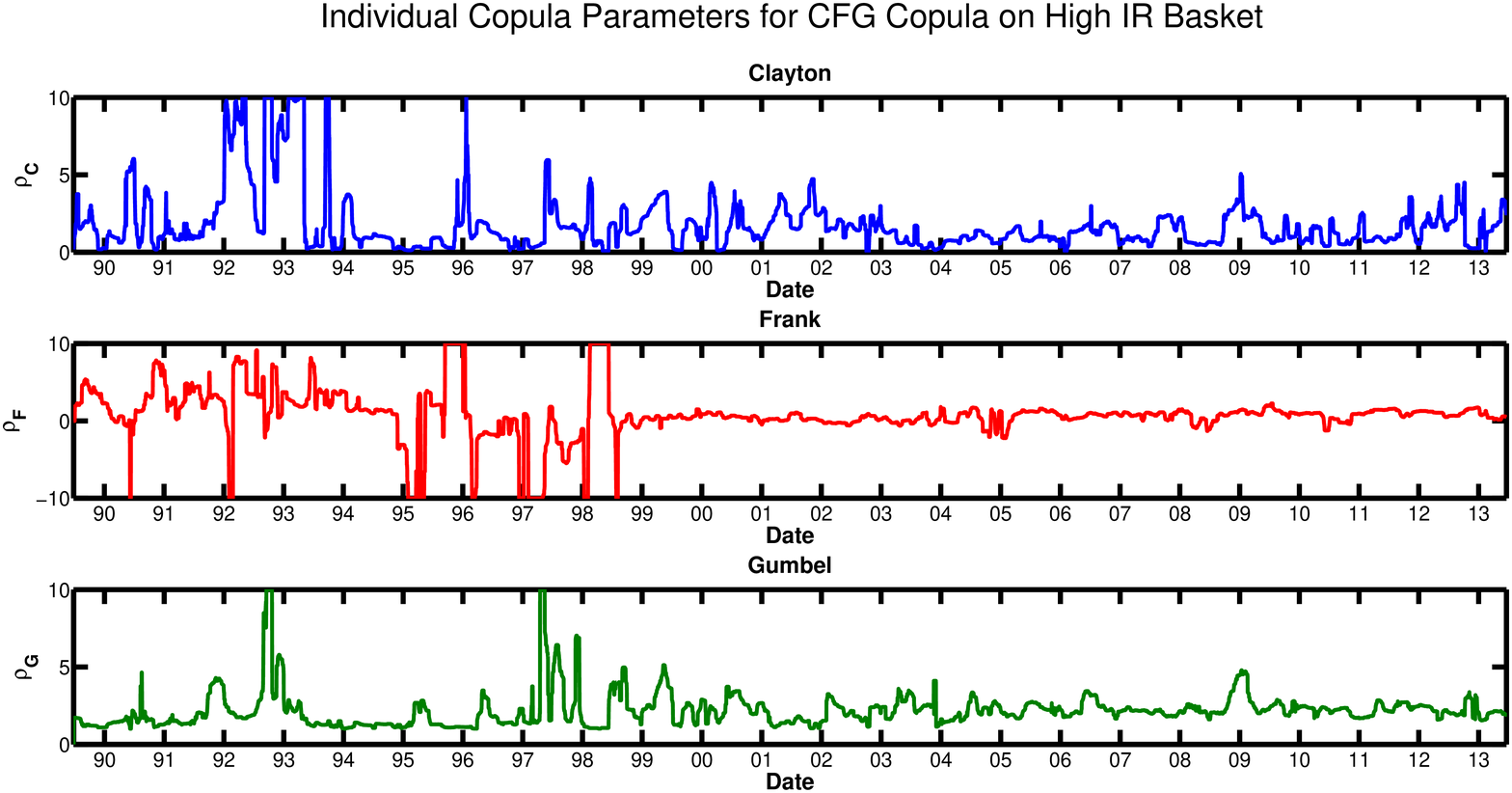}
\caption{$\rho$ Copula parameters for the Clayton, Frank and Gumbel copulae
on the high interest rate basket.}
\label{rho_high_49}
\end{figure}

\begin{figure}[p]
\centering
\includegraphics[width =\textwidth , height=90mm]{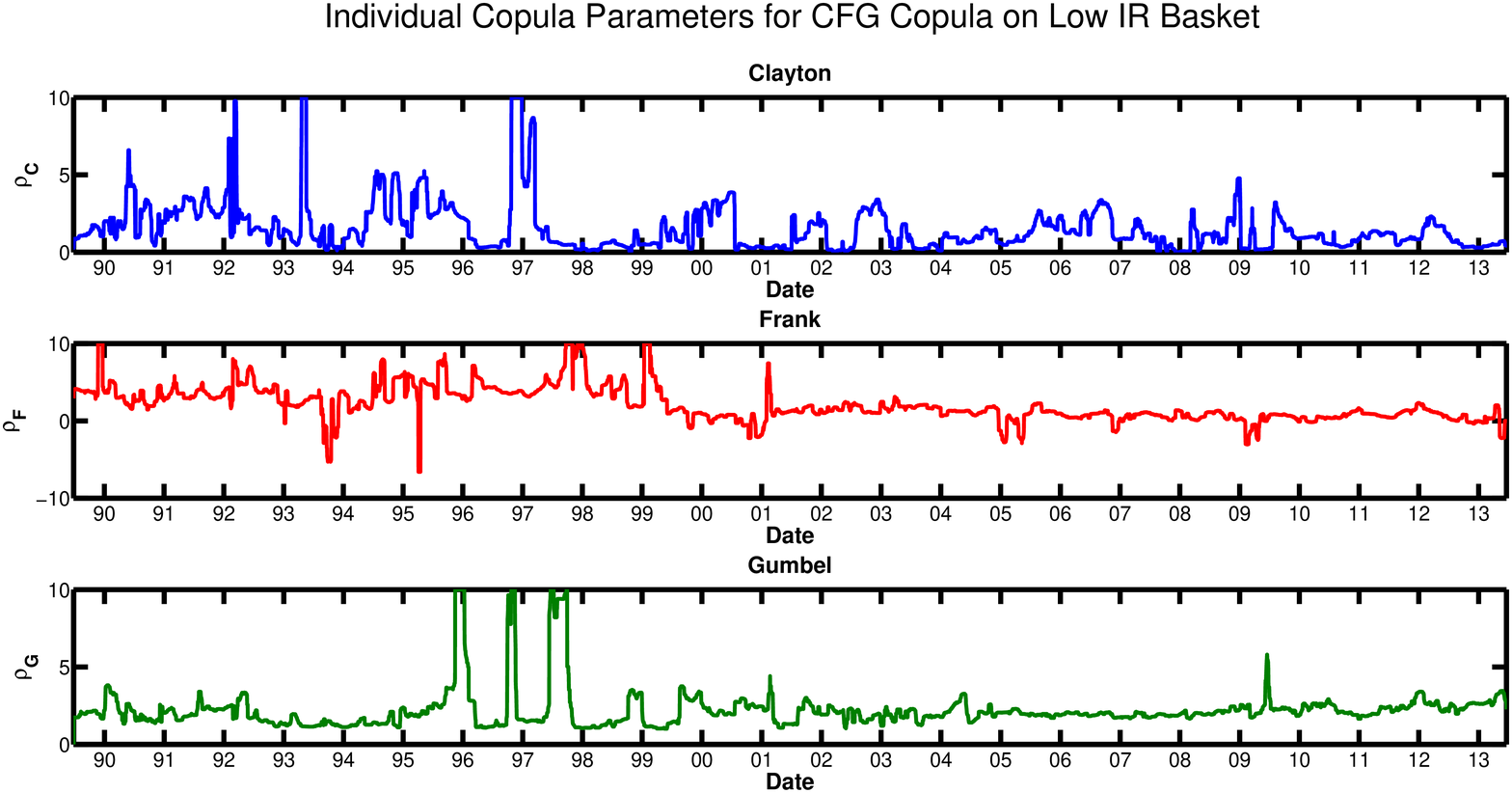}
\caption{$\rho$ Copula parameters for the Clayton, Frank and Gumbel copulae
on the low interest rate basket.}
\label{rho_low_49}
\end{figure}

\begin{figure}[p]
\centering
\includegraphics[width =\textwidth , height=90mm]{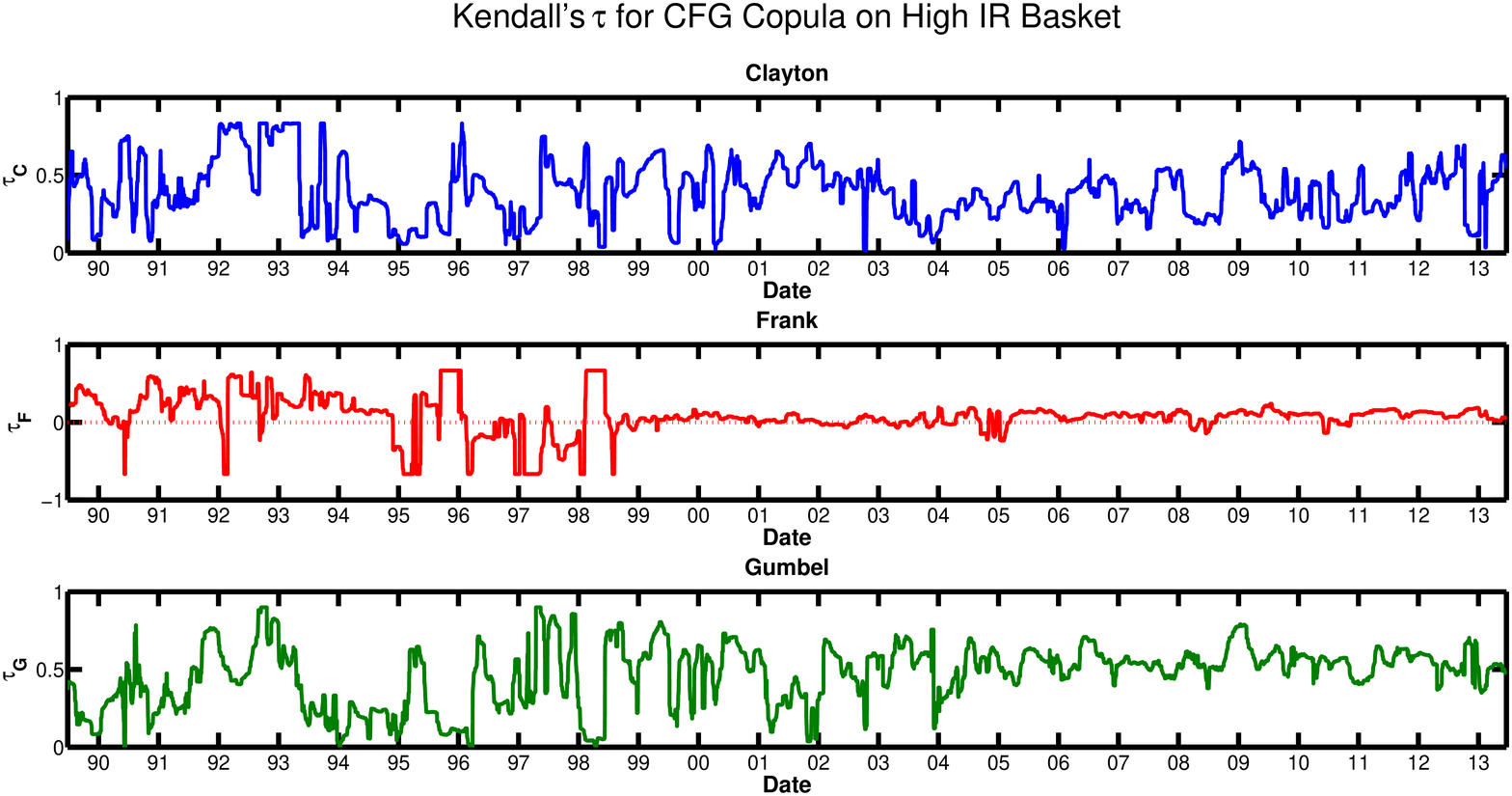}
\caption{Kendall's $\tau$ for the Clayton, Frank and Gumbel copulae
on the high interest rate basket.}
\label{tau_high_49}
\end{figure}


\FloatBarrier

\begin{figure}[h]
\centering
\includegraphics[width =\textwidth , height=90mm]{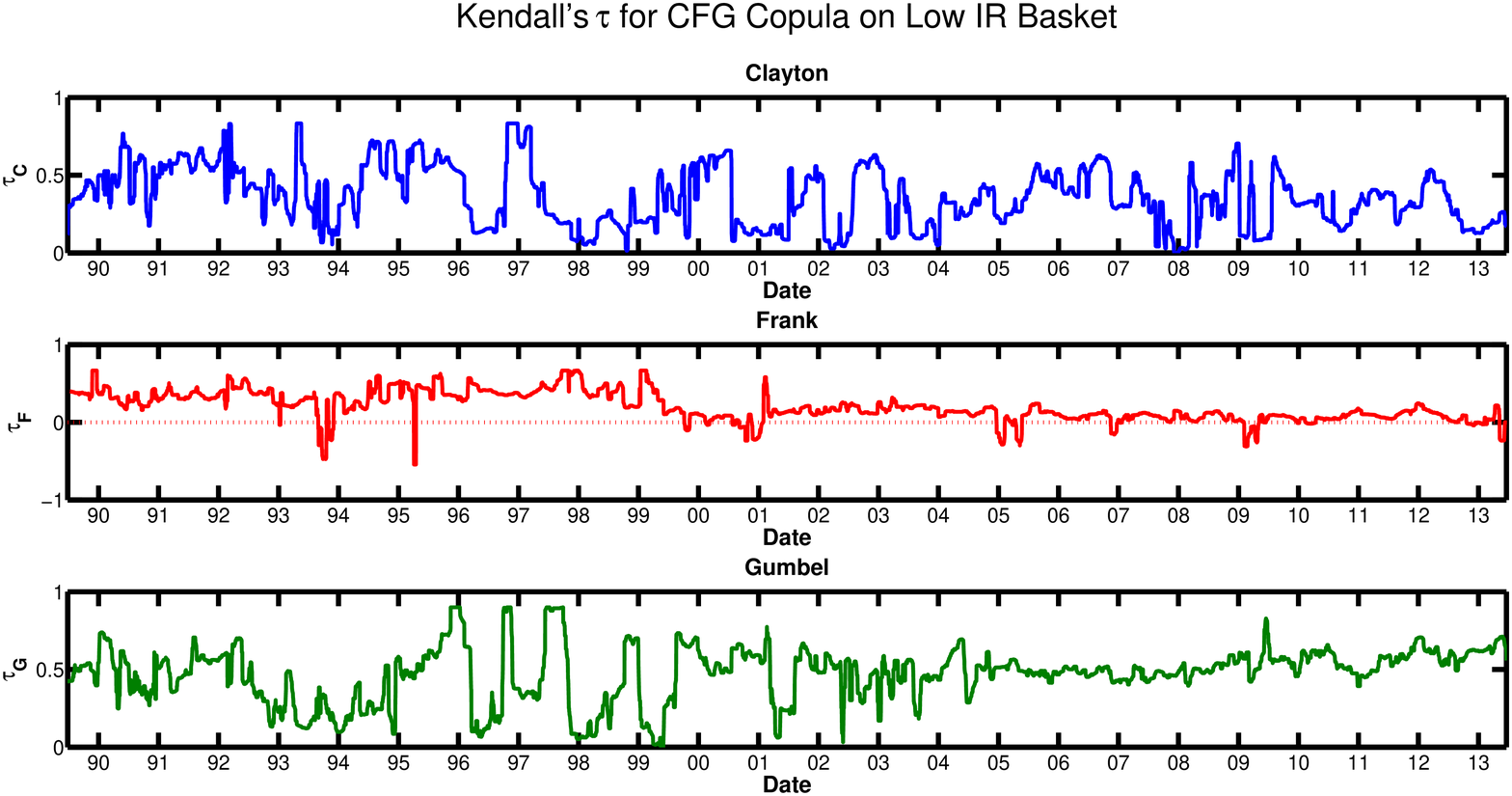}
\caption{Kendall's $\tau$ for the Clayton, Frank and Gumbel copulae
on the low interest rate basket.}
\label{tau_low_49}
\end{figure}

%% file: appendixB_MA_14_Nov_2013_v3_0_0.tex
\section*{Appendix B}

\section*{Clayton-Frank-Gumbel Mixture Copula Density}
\vspace{5mm}
\begin{multline}
\begin{aligned}
c^{CFG}_{\rho_C, \rho_F,  \rho_G}(\mathbf{u}) = & \lambda_C (c^C_{\rho_C}(\mathbf{u})) + \lambda_F (c^F_{\rho_F}(\mathbf{u})) + \lambda_G(c^G_{\rho_G}(\mathbf{u})) \\
= & \lambda_C \times \prod\limits_{k=0}^{d-1}(\rho k +1) \left( \prod\limits_{i=1}^d u_i \right)^{-(1+\rho)} \left(1 + t_\rho^C \left(\mathbf{u} \right) \right)^{\left(-d +\frac{1}{\rho} \right)} \\
& + \lambda_F \times  \left(\frac{\rho}{1 - e^{-\rho}}\right)^{d - 1} Li_{-(d - 1)} \left\{h_\rho^F ({\bf u}) \right\} \frac{e^{\left( -\rho \sum\limits_{j=1}^d u_j  \right)}}{h_\rho^F ({\bf u })} \\
& + \lambda_G \times \rho^d e^{\left(-t_\rho (\mathbf{u})^\frac{1}{\rho}\right)} \frac{\prod\limits_{i=i}^d (-log\; u_i)^{\rho-1}} {t_\rho(\mathbf{u})^d \prod\limits_{i=1}^d u_i} P_{d,\frac{1}{\rho}}^G (t_\rho^G(\mathbf{u})^\frac{1}{\rho})
\end{aligned}
\end{multline}

\noindent where \newline
$h_\rho^F ({\bf u}) = \left(1 - e^{-\rho} \right)^{1 - d} \prod\limits_{j=1}^d \left\{1 - e^{-\rho u_j} \right\}$




%% file: appendixC_MA_06_Dec_2013_v3_0_0.tex


\renewcommand{\arraystretch}{3.2}
\renewcommand{\thefootnote}{\fnsymbol{footnote}}	
\setlength{\tabcolsep}{24pt}				
\setcounter{footnote}{0}	

\begin{landscape}
\begin{table}[htbp]
  \centering
    \caption{Archimedean copula generator functions, inverse generator functions and generator function d-th derivatives.}
    \begin{tabular}{rrrr}
    \toprule
    \textbf{Family} & \boldmath{}\textbf{$\psi$}\unboldmath{} & \boldmath{}\textbf{$\psi^{-1}$}\unboldmath{} & \boldmath{}\textbf{$(-1)^d \psi^{(d)} $}\unboldmath{}  \\
    \midrule
    \textbf{Clayton} & $(1 + t)^{-\frac{1}{\rho}}$ & $(s^{-\rho} - 1)$ & $\frac{\Gamma \left(d+\frac{1}{\rho} \right)}{\Gamma \left(\frac{1}{\rho} \right)} (1+t)^{- \left(d+\frac{1}{\rho} \right)}$ \\
    \textbf{OP-Clayton} & $ \left(1 + t^{\frac{1}{\beta}} \right)^{-\frac{1}{\rho}}$ & $(s^{-\rho} - 1)^{\beta}$ & $ \frac{ \sum_{k=1}^d \tablefootnote{$a_{dk}^G (\frac{1}{\rho}) = \frac{d!}{k!} \sum_{i=1}^k \binom{k}{i} \binom{i/\rho}{d} (-1)^{d-i} \;, \;\; k \in {{1,...,d}}$} a_{dk}^G (\frac{1}{\beta}) \frac{\Gamma \left(k+\frac{1}{\rho} \right)}{\Gamma \left(\frac{1}{\rho} \right)} \left(1+t^{\frac{1}{\beta}} \right)^{- \left(k+\frac{1}{\rho} \right)} \left(t^{\frac{1}{\beta}} \right)^k }{t^d}$ \\
    \textbf{Frank} & $- \frac{1}{\rho} \ln \left[1 - e^{-t}(1 - e^{-\rho}) \right]$ & $-\ln \frac{e^{-s\rho} - 1}{e^{-\rho} - 1}$ & $\frac{1}{\rho} \tablefootnote{$Li_s(z) = \sum_{k=1}^\infty \frac{z^k}{k^s}$} Li_{-(d - 1)}  \{(1 - e^{-\rho}) e^{-t}\}$ \\
%
    \textbf{Gumbel} & $e^{-t^{\frac{1}{\rho}}}$ & $(-\ln s)^\rho$ & $\frac{\psi_\rho(t)}{t^d} \tablefootnote{$P_{d,\frac{1}{\rho}}^G \left(t^{\frac{1}{\rho}} \right) = \sum_{k=1}^d a_{dk}^G \left(\frac{1}{\rho} \right)  (t^\frac{1}{\rho})^k$} P_{d,\frac{1}{\rho}}^G \left(t^\frac{1}{\rho} \right)$ \\
%
    \bottomrule
    \end{tabular}%
  \label{generators}%
\end{table}%
{\bf Remark:} The densities for the one-parameter copulae in Table \ref{generators} can be calculated using Equation \ref{eq:density}. For details of the results contained in this table see \cite{hofert2012}.

\end{landscape}

\renewcommand{\arraystretch}{2}
\setlength{\tabcolsep}{6pt}